\documentclass{amsart}

\usepackage{amsmath}
\usepackage{amssymb}
\usepackage{amsthm}
\usepackage{amsfonts}
\usepackage{amstext}
\usepackage{amsopn}
\usepackage{amsxtra}
\usepackage{graphicx}
\usepackage{mathrsfs}

\newcommand{\E}{\mathrm{e}}
\newcommand{\I}{\mathrm{i}}

\newcommand{\D}{\mathrm{d}}

\newcommand{\be}{\begin{equation}}
\newcommand{\ee}{\end{equation}}

\DeclareMathOperator{\im}{Im}

\DeclareMathOperator{\re}{Re}

\newtheorem{lemma}{Lemma}[section]

\theoremstyle{definition}

\theoremstyle{definition}
\newtheorem{remark}[lemma]{Remark}
\theoremstyle{definition}

{\catcode `\@=11 \global\let\AddToReset=\@addtoreset}
\AddToReset{equation}{section}

\newcommand{\N}{{\mathbb N }}
\newcommand{\R}{{\mathbb R}}
\newcommand{\C}{{\mathbb C}}
\newcommand{\e}{{\epsilon }}
\newcommand{\ie}{{\sl i.e.\/ }}
\newcommand{\cf}{{\sl cf.\/ }}
\newcommand{\eg}{{\sl e.g.\/}}

\def\d{{\partial}}

\def\({\left(}
\def\){\right)}
\def\<{\left\langle}
\def\>{\right\rangle}
\def\O{\mathcal O}

\renewcommand{\L}{{L}}
\newcommand{\Id}[1]{{\rm 1\kern-2pt I_{#1}}}
\renewcommand{\hbar}{{\displaystyle\bar{\phantom{x}}\kern-6pt h}}

\numberwithin{equation}{section}

\begin{document}


\title[Kadomtsev-Petviashvili equation]{Numerical study of oscillatory regimes in the Kadomtsev-Petviashvili equation}

\author[C. Klein]{Christian Klein}
 \address{Max Planck Institute for Mathematics in the Sciences} 
 \email{klein@mis.mpg.de}

\author[C. Sparber]{Christof Sparber}
 \address{Wolfgang Pauli Institute Vienna \& Faculty of Mathematics, Vienna University, Nordbergstra\ss e 15, A-1090 Vienna, Austria}
\email{christof.sparber@univie.ac.at}

\author[P. Markowich]{Peter Markowich}
 \address{Faculty of Mathematics, Vienna University, Nordbergstra\ss e 15, A-1090 Vienna, Austria}
\email{peter.markowich@univie.ac.at}

\begin{abstract}
The aim of this paper is the accurate numerical study of the KP equation. In particular we are concerned with 
the small dispersion limit of this model, where no comprehensive 
analytical description exists so far. To this end we first study a similar
highly oscillatory regime for 
asymptotically small solutions, which can be described via the Davey-Stewartson system. In a second step we investigate 
numerically the small dispersion limit of the KP model in the case of large amplitudes. Similarities and differences 
to the much better studied Korteweg-de Vries situation are discussed as well as the 
dependence of the limit on the additional transverse coordinate.
\end{abstract}
\thanks{We thank B.~Dubrovin, E.~Ferapontov, 
J.~Frauendiener, and T.~Grava for 
helpful discussions and hints. 
This work has been supported by the Wittgenstein Award 2000 of the second author. 
C.S. has been supported by the APART research grant funded by
the Austrian Academy of Sciences.}
\maketitle

\begin{center}

version: \today

\end{center}

\section{Introduction}\label{sint}

This work is concerned with the $2+1$ dimensional \emph{Kadomtsev-Petviashvili equation} (KP), given by
\be
\partial_x {\left(\partial_t u + u \, \partial_x u +  \e^2 \partial_{xxx} u \right)} + \lambda \, \partial_{yy} u =0, \quad \lambda = \pm 1,
\label{KP}
\ee
where $(t,x,y)\in \R_t\times \R_x\times \R_y$ and where $\e >0$ is a (small) scaling parameter, as introduced below.
The case $\lambda = -1$ corresponds to the so-called KP-I model, whereas $\lambda = + 1$ is usually referred 
to as KP-II equation. Both cases can be derived as models for nonlinear dispersive 
waves on the surface of fluids \cite{KaPe} (see also \cite{Jo}). Thereby it is assumed that the propagation of the waves 
is essentially one-dimensional with weak transverse effects. Roughly speaking, 
KP-I describes the case when surface tension is strong, whereas KP-II is a good model for weak surface tension. 
Moreover, the KP type equations also arise as a model for sound waves in ferromagnetic media \cite{TuFa} and in the 
description of two-dimensional nonlinear matter-wave pulses in Bose-Einstein condensates, see \eg \ \cite{HMV, JoRo}.

Notice that the KP equation has originally been derived in the following form
\be\label{KPor}
\begin{split}
\partial_t u + u \, \partial_x u + \e^2 \partial_{xxx} u+ \lambda \, \partial_{y} v =& \ 0,\quad u\big |_{t=0} = u_{\rm I} (x,y),\\ 
\partial_y u = & \  \partial_x v,
\end{split}
\ee
where the second equation can now be seen as a constraint for the Cauchy problem given in the first line. Therefore 
we shall also impose for equation \eqref{KP} initial data in the Schwartz class of smooth and rapidly decreasing functions, \ie  
\be \label{initialdata}
u\big |_{t=0} = u_{\rm I} \in \mathcal S(\R_x\times \R_y),
\ee
even though an interpretation of \eqref{KP} in terms of a Cauchy problem seems not completely obvious at first sight, 
due to the appearing mixed derivative $\partial_{xt}u$. 
Nevertheless this is the usual 
approach when dealing with \eqref{KP} and it is strengthened by the 
fact that  in large parts of the literature 
one considers the KP model in its so-called \emph{evolutionary form} 
\be\label{KPnew}
\partial_t u + u \, \partial_x u + \e^2\partial_{xxx} u+ \lambda \, \partial^{-1}_x\partial_{yy} u = 0,
\quad u\big |_{t=0} = u_{\rm I} (x,y).\\ 
\ee
Here, the anti-derivative $\partial_x^{-1}$ can be uniquely defined via
\be 
\label{invers}
\partial_x^{-1} f(x):= \, \frac{1}{2}\left( \int_{-\infty}^{x} f(\zeta)\, \D \zeta - \int_{x}^{+\infty} f(\zeta)\,  \D \zeta \right),
\ee
at least if $f(x)$ decays sufficiently fast to zero, as $ x \to \pm \infty$, respectively. 
Alternatively $\partial_x^{-1}$ can be seen as a \emph{Fourier multiplier} with the singular symbol $-\I /k_x$, 
as we shall do in the following. 

In this paper we are interested in the accurate numerical simulation of the KP equation 
using a spectral scheme with preconditioning. 
To this end we remark that alternative numerical schemes for $2+1$ dimensional wave equations 
have been proposed in \cite{Dr, FKM}.
Moreover, earlier numerical studies of the KP equation can also be found 
in, \eg, \, \cite{ISS, KaEl, SeIn, WAS, Wa}, aiming mostly at the description of (interacting) solitons. 
In our work though the focus will be mainly on asymptotic regimes of the KP model. In particular we are  
concerned with the so-called \emph{dispersionless limit}, \ie the limit of \eqref{KP} as $\e\rightarrow 0$. 

To motivate this study consider the unscaled (dimensionless) KP model 
\be
\partial_{X} 
{\left(\partial_{T} u + u \, \partial_{ X} u +   \partial_{ X X X} u \right)} + \lambda \, \partial_{ Y Y} u =0, 
\quad \quad u\big |_{T=0} = u_{\rm I} ( X, Y), 
\label{unKP}
\ee
where $(T,  X,  Y)$ are now considered as the natural scales of observation, or microscopic scales.
In \eqref{unKP} we shall now introduce \emph{slowly varying} solutions of the form, \ie
\be\label{ssol}
u=u(\e T, \e  X, \e  Y), \quad 0<\e\ll 1,
\ee
where $\e$ is a small scaling parameter, \ie the \emph{microscopic/macroscopic scales ratio}. 
These slow variations, for example, can be introduced via a class of corresponding initial data. 
Plugging \eqref{ssol} into \eqref{unKP} and denoting $(t,x,y) = (\e T, \e  X, \e Y)$, 
we obtain the scaled equation 
\eqref{KP} with $\e \ll 1$. 
In the following we shall always consider the KP equation in the form (\ref{KP}). 
The notion of the coordinates thus refers to the macroscopic sales.

In the \emph{formal limit} $\e \to 0$ we get the \emph{dispersionless KP equation} (dKP), \ie  
\be\label{dKP}
\partial_x {\left(\partial_t u + u \, \partial_x u \right)} + \lambda \, \partial_{yy} u =0.
\ee
Clearly this is a singular limiting procedure, which requires particular care.
The analogous problem for the scaled \emph{Korteweg-de Vries equation} (KdV), see equation 
\eqref{KdV} below, has been intensively studied by several authors, \cf \cite{GT, 
LL, Le, MS}. More recently this problem has also been treated numerically in \cite{GrKl}. 
In the KdV case, it is known that the solution of the Cauchy problem for smooth hump-like initial data is characterized by 
the appearance of a zone of \emph{rapid modulated oscillations} \cite{GrKl, FRT1}. A 
rigorous asymptotic treatment of these oscillations has been established, 
relying heavily on inverse scattering techniques and complete integrability. 
On the other hand, a complete mathematical description of the 
small dispersion limit for the KP model has not yet been achieved \cite{Bo, Kr}. 
The purpose of this paper is thus to explore this problem 
numerically. Similar to KdV, it is expected that the dKP model will develop \emph{shocks} in 
finite time. In this case, the formal limit \eqref{unKP} is no longer an accurate description of \eqref{KP} as $\e\to 0$. 
For small, but still non-zero $\e$, the dispersive term $\propto\d_{xxx}u$ in (\ref{KPnew}) 
presumably will smooth out these shocks by forming an oscillatory zone. 
This work provides numerically evidence for these facts.

More precisely the \emph{outline of this paper} is as follows: In the next section we shall first recall some well known features of 
the KP equation. Also, we shall derive the Green's function of the corresponding linear model, which already 
gives some insight on the oscillatory nature of the considered equation. In Section \ref{salg} the numerical algorithm is 
presented and we shall test its performance for two cases of soliton propagation. In Section \ref{smod} we discuss the the asymptotic 
behavior for $\e\ll 1$ in the particular case of small solutions, \ie with amplitudes of order $\O(\e)$. This yields an asymptotic description 
via the Davey-Stewartson system, which is studied numerically for the case $\lambda = 1$. 
The corresponding asymptotic error to the KP-II equation, defined in the $L^2(\R^2)$ sense, 
is, within numerical precision, found to be $\O(\e^{5/2})$ which is remarkable since it is exactly 
as expected from the corresponding rigorous KdV studies in \cite{Sch}. Finally we turn to the problem of small 
dispersion for solutions of order $\O(1)$ in Section \ref{slow}. We 
shall numerically investigate the entropy solution of the dKP equation and the corresponding 
oscillatory behavior in the KP model, in particular its dependence on the $y$-coordinate. 
We also find numerically that the asymptotic $L^2(\R^2)$-error in this case is $\O(\e^{3/2})$, at least 
before the appearance of the first shock in the corresponding dKP solution.


\section{Properties of the KP equation} \label{slin}

\subsection{Preliminaries}
In the context of water waves, the KP equation can be seen as a two-dimensional generalization of the 
celebrated KdV equation \cite{Jo}, \ie
\be\label{KdV}
\partial_t u + u \, \partial_x u + \e^2\partial_{xxx} u=0,\quad u\big |_{t=0} = u_{\rm I} (x).
\ee
Clearly, any solution to KdV is a $y$-independent solution to the KP equation, forming the \emph{KdV sector} of \eqref{KP}. 
Indeed similar to KdV, the KP model belongs to the class of \emph{integrable systems} 
and is known to conserve an infinite number of quantities, among them \emph{mass}
\be \label{mass}
\mathrm{M}[u(t)]:=\int_{\R^2} u^2(t,x,y) \, \D x  \, \D y 
\ee 
and \emph{energy} 
\be \label{energy}
\mathrm{E}[u(t)]:= \frac{1}{2} \int_{\R^2} \left(\partial_x u(t,x,y))^2 - \lambda 
(\partial_x^{-1} \partial_y u(t,x,y))^2 - \frac{\e^2}{3} \, u^3(t,x,y) 
\right)
\D x \, \D y ,
\ee
as well as momentum etc., \cf \cite{AbCl} for more details. 
Again similar to KdV, explicit solutions to the KP model can be obtained via \emph{inverse scattering} methods. 
In the following though we shall not go into more details on this important topic, arising in the study of integrable systems, 
and rather refer to, \eg, \cite{APS, NMPZ} and the references given therein. 

Despite the apparent similarity of KP-I and KP-II, the two equations behave rather differently 
both qualitatively and quantitatively.  
For example, only the KP-I has the advantage of having an energy which is positive definite in 
the leading order terms. 
Additionally, unlike the KP-I model, the KP-II equation does \emph{not} 
admit so-called \emph{lump solitons}. Generally speaking, a KP soliton is a solution of the form
\be
u(t,x,y)=u(x-t v_{x} ,y- t v_{y} ),
\ee
with (constant) \emph{group velocity} $v_{\rm g}=(v_x,v_y) \in \R^2$. Clearly, any KdV soliton is also a KP soliton solution. 
We also remark that both KP models bear several differences in what 
concerns the stability or instability of solitons, \cf \cite{APS, ISR} and 
the references given therein. 
The single lump soliton to \eqref{KP}, with $\lambda = -1$, has the following form (here, we set $\epsilon =1$ for simplicity)
\begin{equation}
    u(t,x,y) = \frac{ 24\,c\, (1-c (x-3ct )^{2}+3c^{2}y^{2})}{(
    1+c (x-3c t)^{2}+3c^{2}
    y^{2})^{2}}
    \label{lump},\quad c \in \R.
\end{equation}
Note that the lump soliton only decays algebraically in both spatial 
directions, as $|x|,|y|\to \infty$. On the other hand both equations admit 
so-called \emph{line soliton} solutions. 
These are solitons which are infinitely extended in 
one spatial direction and which exponentially decay in 
the other direction. The simplest solution of this kind is a
$y$-\emph{independent} 1-soliton of the KdV equation (again $\e =1$ for simplicity) 
\begin{equation}
    u(t,x)= \frac{12}{\cosh^{2}(x-x_{0}-4t)}\, , \quad \mbox{with $x_0\in \R$ fix}.
    \label{linesol}
\end{equation}

Solutions which are even better localized in $x$ and $y$ cannot be expected (see also \cite{BoMa}). 
Indeed it is a common feature of both KP-I and KP-II that, even if the initial data is a Schwartz function 
the corresponding solutions $u(t)$ for $t>0$ in general will \emph{not} 
stay in the Schwartz class, unless $u_{\rm I}$ satisfies in addition an infinite number of constraints. 
Rather it is known that, for generic initial data, the solutions $u(t)$ will develop ``tails'' 
decaying only algebraically in certain spatial directions. 
The reason for this behavior is already seen on the level of the Green's function for the linear KP 
equation, as we shall discuss in the next subsection. 

\subsection{The linear KP equation} \label{sub} 
In the following we shall study the linear part of the KP equation, \ie 
\be\label{linKP}
\partial_x {\left(\partial_t u +  \e^2 \partial_{xxx} u \right)} + \lambda \, \partial_{yy} u =0, 
\quad u\big|_{t=0}=u_{\rm I}(x,y).
\ee
for smooth (finite mass) solutions $u(t)\in L^2 (\R^2)\cap L^1(\R^2)$. If we denote by
$$
(\mathcal F u(t))(k_x,k_y)\equiv \widehat{u}(t,k_x,k_y):=\int_{\R^2} u(t,x,y) \, \E^{-\I (k_x x + k_y y)}\, \D x \, \D y,
$$
the \emph{Fourier transform} of $u(t)$ w.r.t $x$ and $y$, the above given equation \eqref{linKP} is equivalent to 
\begin{equation}
\I k_x \partial_t \, \widehat{u} +\epsilon^{2} k_{x}^{4}\, \widehat{u}- \lambda k_{y}^{2} \, \widehat u=0.
\end{equation}
In order to pass to the evolution form \eqref{KPnew} we need to be able to apply 
the singular multiplier $-\I /k_x$ to this equation, to obtain 
\begin{equation}
\label{flinKP}
    \partial_t \widehat{u}+ \I \left( \frac{\lambda k_{y}^{2}}{k_{x}}-\epsilon^{2}  k_{x}^{3}\right)\widehat{u}=0.
\end{equation}
Having in mind that $\widehat u(t)\in C_{\rm b}(\R^2)$, for $u(t)\in L^1(\R^2)$, this certainly requires that 
$\widehat u(t,0,k_y)=0$, for all $k_y\in \R$.  
To ensure this, we shall impose the following condition on the initial data
\be \label{const}
\int_{\R}  u_{\rm I}(x,y)\, \D  x  =0,  
\ee
which, in Fourier space, is equivalent to $u_{\rm I}(0,k_y)=0$, for all $k_y \in \R$. 
In this case the (weak) solution of \eqref{flinKP} is obviously given by
\be \label{linsol}
\widehat u(t, k_x,k_y) = \E^{-\I t(\lambda k_{y}^{2}/k_{x} -\epsilon^{2} k_{x}^{3})} u_{\rm I} (k_x, k_y),
\ee
which satisfies $\widehat u(t,0,k_y)=0$, for all $t\in \R$.
The constraint \eqref{const} thus seems to be quite natural when passing to the formulation \eqref{KPnew}, 
\cf \cite{BPP, Tz}. Moreover it can also be seen from the numerical example given in Section \ref{salg}, 
that initial data which do not satisfy \eqref{const} become non-differentiable for any $t\not =0$.
\begin{remark} 
Alternatively one might also remark that (formally) integrating \eqref{KP} w.r.t. to $x \in \R$ 
gives 
\be
\int_{\R}  \partial_{yy} u(t, x,y)\, \D  x  =0,\quad \forall \, y \in \R, t\not =0,
\ee
or, equivalently, $-k_y^2 \, \widehat u(t,0,k_y)=0$, 
a condition certainly fulfilled by \eqref{linsol}, \eqref{const}. 
\end{remark}
Of course $\widehat u(t,0,k_y)=0$ is not sufficient to obtain a strong KP-solution (being differentiable in-time) 
which satisfies \eqref{flinKP} point wise in Fourier space. To this end one would require a sufficient fast decay of 
$\widehat u(t,k_x,k_y)$, as $k_x \to 0$. Therefore most of the literature is concerned with the mild 
formulation \eqref{linsol}, which is consequently translated to the non-linear model, via Duhamel's formula
\be
u(t,x,y)=  U(t) \, u_{\rm I} (x, y) - 
\int_{0}^t U(t-s) \, u(s,x,y) \partial_x u(s,x,y) \, \D s.
\ee
Here, $U(t)$ denotes the unitary group in $L^2(\R^2)$ 
defined via its symbol in Fourier space, \ie
\be
\widehat U(t):=\exp\left(-\I t(\lambda k_{y}^{2}/k_{x} -\epsilon^{2} k_{x}^{3})\right).
\ee 
Following this approach, well posedness issues are usually studied in non-isotropic Sobolev 
spaces $H^{s_1}(\R_x)\times H^{s_2}(\R_y)$. 
It should be noted that local and global well posedness results are much more complete in the 
KP-II case than in the case of KP-I. The first result in this direction has been 
obtained in \cite{Bou} proving, that the KP-II model is \emph{globally well posed} for 
$u_{\rm I} \in \L^2(\R^2)$, or $u_{\rm I} \in \L^2(\mathbb T^2)$. As 
far as Sobolev regularity is concerned 
we shall only remark that the required index pair for KP-II is: $s_1> -1/3$, $s_2\geq 0$ \cite{Ta}, 
referring to \cite{CKS, MST, TaTz, Tz} for more details on these issues.
\begin{remark}
Note that the formal identification $\partial_{tx} u = \partial_{xt} u$ requires particular care 
if one does not consider solutions, which are differentiable in time \cite{MST}. 
In this context we also remark that sometimes even weaker notions of solutions 
which are only defined to be distributional in-time, are taken into account, \cf \cite{BPP}. 
\end{remark}

In order to get more insight on the dynamics of the linear equation we first note that \eqref{linsol} implies 
\be \label{udarst}
u(t, x,y) = \frac{1}{4\pi^2} \left(\mathcal F^{-1}\widehat U(t)\right) \ast u_{\rm I} (x, y),
\ee
where $\ast$ denotes the convolution w.r.t. $x\in \R$ and $y\in \R$.
In order to calculate the inverse Fourier transform of $\widehat U(t)$ we shall split the transformation in two parts via
\be \label{invU}
\mathcal F^{-1}\left(\E^{-\I t(\lambda k_{y}^{2}/k_{x} -\epsilon^{2} k_{x}^{3})}\right) = 
\mathcal F^{-1} \left(\E^{-\I t\lambda k_{y}^{2}/k_{x}}\right) \ast \mathcal F^{-1}\left( \E^{\I t \epsilon^{2} k_{x}^{3}}\right).
\ee
In the following let us restrict to the case of solutions for times $t>0$, for simplicity. Next, 
consider the last factor on the r.h.s. of \eqref{invU} and recall the Fourier representation of the so-called 
\emph{Airy function}, \ie 
\be 
{\mathrm {Ai}}(x)= \frac{1}{2\pi}\int_\R \E^{\I (x k + k^3/3)} \D k.
\ee  
It is well known, see, \eg, \cite{AbSt}, that the function ${\mathrm {Ai}}(x)$ admits a series expansion in the form 
\be
{\mathrm {Ai}}(x)= \frac{1}{3^{2/3}\pi} \sum_{n=0}^\infty \frac{\Gamma\left(\frac{n+1}{3}\right)}{n!} 
\left(3^{1/3} x\right)^n \sin\left(\frac{2(n+1)\pi}{3} \right).
\ee
Moreover ${\mathrm {Ai}}(x)$ obeys two completely different asymptotic limits as $x\to \pm \infty$, respectively, namely
\be \label{airyas}
{\mathrm {Ai}}(x)\stackrel{x\to +\infty}\sim \frac{1}{2 \pi^{1/2} x^{1/4}}\, \E^{-\frac{x^{3/2}}{3}}, 
\quad {\mathrm {Ai}}(x)\stackrel{x\to -\infty}\sim \frac{1}{\pi^{1/2} x^{1/4}}\, \cos\left(\frac{2x^{3/2}}{3}  - \frac{\pi}{4} \right).
\ee
We consequently obtain, by substituting $k= k_x(3t\e^2)^{1/3}$, that 
\be \label{inv1}
\mathcal F^{-1}\left( \E^{\I t \epsilon^{2} k_{x}^{3}}\right) = \frac{1}{(3t\e^2)^{1/3}}\,
{\mathrm {Ai}}\left(\frac{x}{(3t\e^2)^{1/3}}\right) \otimes \delta(y), \quad \forall \, t>0,
\ee
in $\mathcal S'(\R_x\times \R_y)$, \ie the sense of tempered distributions. 

On the other hand, computing the inverse Fourier transform of the first factor on the r.h.s. 
of \eqref{invU} yields, after some lengthy computations, that (for $t>0$)
\be \label{inv2}
\mathcal F^{-1} \left(\E^{-\I t\lambda k_{y}^{2}/k_{x}}\right) = 
\left\{ \begin{aligned}&\,  \frac{4t^{1/2}}{(4xt+\lambda y^2)^{3/2}}, \quad \mbox{if: $4xt + \lambda y^2>0$,} \\
&\ 0 \quad \quad \quad \quad \quad \quad \quad \  \mbox {else.}
\end{aligned}
\right.
\ee
Again this has to be interpreted as a distribution on Schwartz space. 
(We note that formula \eqref{inv2} is equivalent to the one found in \cite{BPP}, 
where only the KP-I case is considered though.) 
Expression \eqref{inv2} yields the Green's function for the linear partial differential operator
\begin{equation}
    L f:=\left(\partial_{xt}+\lambda \partial_{yy}\right)f=0, \quad \lambda=\pm1
    \label{dkpl}.
\end{equation}
Note that $Lf =0$ is a hyperbolic PDE as can be seen by introducing the coordinates 
$\alpha =x+t$ and $\beta =x-t$, and transforming (\ref{dkpl}) into 
\begin{equation}
    \left(\partial_{\alpha \alpha }-\partial_{\beta \beta }+\lambda \partial_{yy}\right)f =0, \quad \lambda=\pm1
    \label{dkpl2}.
\end{equation}
This is nothing but the standard $2+1$ dimensional \emph{wave equation}, where $\alpha$ or $\beta$ take on the role of a 
``time variable'' for $\lambda=-1$ and $\lambda=1$, respectively. 
Since the operator $L$ can also be interpreted as the linear part of the dKP equation \eqref{dKP}, 
imposing at time $t=0$ an initial condition of the form \eqref{initialdata} indeed 
furnishes a \emph{characteristic Cauchy problem}. Moreover, the space-time region determined by $4xt + \lambda y^2=0$ 
is also a characteristic which explains the divergences of the Green's function \eqref{inv2} there.

Combining \eqref{udarst} with the formulas \eqref{inv1} and \eqref{inv2} 
we can draw several important consequences:
\begin{itemize}
\item The $y$-dependence of the solutions to the linear KP equation \eqref{linKP} 
is completely non-oscillatory and has a rather slow algebraic decay as $|y|\to \infty$.
\item For any time $t>0$ we obtain infinitely extended tails of the solution within 
the region determined by $4xt + \lambda y^2>0$, for $t>0$. 
\item The oscillatory behavior of the linear KP equation is governed by the Airy function. 
In our scaling this yields oscillations with wave length $\O(\e)$ in the $x$-direction, as can be seen from 
\eqref{airyas}.  
\end{itemize}


\section{The numerical algorithm} \label{salg}

\subsection{A Spectral approach with preconditioning} 
We are interested in the numerical solution of the KP equation for rapidly decreasing, 
smooth initial data $u_{\rm I}$. 
To study the zone of fast oscillations, due to $\epsilon \ll 1$, 
it is convenient to use a periodic setting of sufficiently large period. This allows for the 
use of spectral methods which are of \emph{high accuracy} and \emph{efficiency} 
because of their excellent approximation properties for smooth 
functions and the existence of fast algorithms for the Fourier 
transformation. The power law tails of the solution, as encountered in the preceding section, 
however will lead to an increasing number of ``\emph{echoes}'' as time goes on, 
due to the chosen periodic domain of computation. For sufficiently large periods though these echoes 
will be small on the studied time-scales and they will \emph{not} influence the 
analysis of the KP oscillations. 

We use here an adapted version of Trefethen's code for the KdV equation 
(Chap.~10 in \cite{trefethen1}) which is available at 
\cite{trefethenweb}. 
The basic idea of the code is the use of a discrete Fourier 
transform in the spatial coordinates as well as of an \emph{integrating factor} such that the time 
derivative is the only linear term appearing in the equation. This 
preprocessing, as introduced in \cite{Hou}, leads to an equation 
without a stiff part and allows for larger time steps. 
The method is slightly more efficient than 
the semi-implicit approach of \cite{FKM}, or time splitting 
techniques such as \cite{HaTa} \eg. More importantly, however, it allows for the 
use of higher order time discretizations which are convenient in the 
context of strong gradients, as needed in Subsection \ref{subsdKP} below. 
Keep in mind that an $n$-th order approach introduces a numerical dispersion of the order 
$(\Delta t)^{n+1}$, which leads to a ``pollution'' of the high spatial frequencies 
in the numerical time evolution. 
On the other hand though we need these high frequencies to resolve numerically the 
solution near a gradient catastrophe in the solution of the dKP model. Thus 
it is helpful to be able to apply a higher order method in the time 
discretization within an efficient approach which allows to deal with 
$2+1$ dimensional settings. 

As before, let $\widehat u(t)$ be the Fourier transform of $u(t)$ 
and denote by $\widehat{u^{2}}$ the Fourier transform of $u^{2}$. Then the Fourier transformed KP 
equation (\ref{KP}) reads
\begin{equation}
    \partial_t \widehat{u}+\left(\frac{\I \lambda k_{y}^{2}}{k_{x}+ \I 
    \lambda 0}
    -\epsilon^{2} \I k_{x}^{3}\right)\widehat{u}+\frac{\I}{2} \,  k_{x}\widehat{u^{2}}=0
    \label{KdVfourier}.
\end{equation}
Equation (\ref{KdVfourier}) has to be  regularized for $k_{x}=0$ in order 
to give numerically sense to the term $-\I/k_x$. Obviously division 
by zero is not allowed, even if the result would make sense 
analytically in a certain 
limiting procedure. To carry out such a limit numerically 
one has to use appropriately chosen finite numbers. 
Here this is done by adding to $k_{x}$ in the denominator a small 
imaginary part of appropriate sign (corresponding to $\lambda$). In the numerics we add the 
smallest floating point number which Matlab can represent, $2.2\ldots 
10^{-16}$. 
Equation (\ref{KdVfourier}) is then equivalent to
\begin{equation}
    \partial_t\left(\E^{\I t(\lambda k_{y}^{2}/(k_{x}+ \I \lambda 
    0)-\epsilon^{2} k_{x}^{3})}
    \widehat{u}\right)+\frac{\I k_{x}}{2} \,  \E^{\I 
    t(\lambda k_{y}^{2}/(k_{x}+\I \lambda 0)- \epsilon^{2}k_{x}^{3})}
    \, \widehat{u^{2}}= 0
    \label{KdVint}.
\end{equation}
To solve equation (\ref{KdVint}) 
numerically we  use the \emph{Fast Fourier Transform} (FFT) in MATLAB for the dependence on the spatial 
coordinates and a fourth-order Runge-Kutta method for the time 
integration for the reasons given above. 
The important term in this integration is $\E^{-\I \Delta t(\lambda k_{y}^{2}/(k_{x}+ \I \lambda 
    0)-\epsilon^{2} k_{x}^{3})}$ as can be seen, \eg, from the simple 
    time discretization
\be\label{tdis}
\widehat u(t+ \Delta t ) = \E^{-\I \Delta t(\lambda k_{y}^{2}/(k_{x}+ \I \lambda 
    0)-\epsilon^{2} k_{x}^{3})} \left(\widehat u(t) - \frac{\I \Delta t k_x}{2} 
    \, \widehat{u^{2}}\right).
\ee
Since we use an explicit method for time integration, stability is an 
issue. We find that due to the integrating factor, time steps of the 
order $\Delta t\sim 1/(N_{x}N_{y})$ lead to a stable time evolution. 
In the following, the empirically found values for the time steps needed in the 
respective computation are always given.

\begin{figure}[!htb]
     \centering \includegraphics[width=10cm]{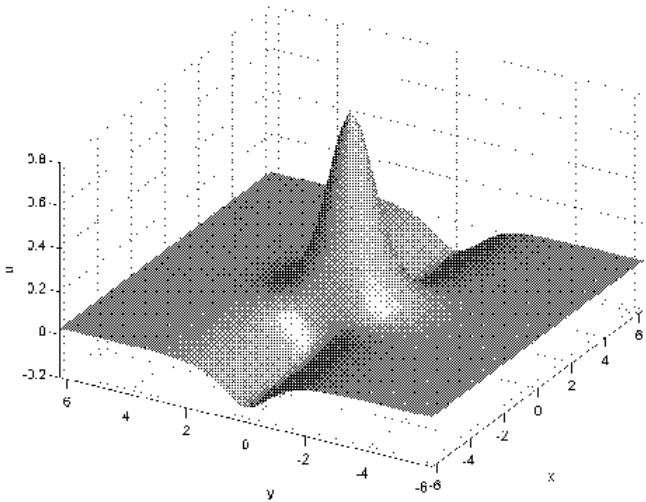}
    \caption{Solution at time $t=4.8\times 10^{-4}$ to the KP-I solution with initial data 
    $1/\cosh^{2}(\sqrt{x^{2}+y^{2}})$ which are not subject to 
    the constraint (\ref{inconst}).}
    \label{figkpns}
\end{figure}

As already discussed in Section \ref{slin}, we require our initial data to be subject to the constraint \eqref{const}. 
In the periodic setting we analogously impose  
\begin{equation}\label{inconst}
    \int_{-\pi L_{x}}^{\pi L_{x}}u_{\rm I}(x,y)\, \D x=0,
\end{equation}
where $2\pi L_{x}$ is the period in $x$. For such data we solve numerically 
for the function $\hat{v}$ defined by $\hat{u}=ik_{x}\hat{v}$ to 
reduce numerical errors.
\begin{remark}
The numerical code is also able to propagate initial data which do \emph{not} satisfy the constraint \eqref{inconst}, 
which however yields a non-smooth solution in arbitrary short times. 
The reason for this is the analytical behavior of the term 
$\E^{-\I t\lambda k_{y}^{2}/(k_{x}+\I\lambda0)}$ appearing in \eqref{tdis}. 
Indeed, for $k_{x}=0$, this term is numerically equal 
to $ 0$, unless  $k_{y}=0$ where it is equal to 1. 
This leads to a continuous but \emph{not} 
differentiable solution at $x=0$, see 
Fig.~\ref{figkpns}. 
\end{remark}

\subsection{Numerical test cases on solitons} To test the accuracy of the code we compare, for the different time 
steps, the numerical solution to an explicitly known exact solution of the KP equation. 
Since the KP equation is completely integrable, large classes 
of explicit solutions are known. The most discussed cases certainly 
are the above-mentioned solitons.
We use here the 1-soliton (\ref{linesol}). 
The propagation of this solution formally does not 
test the $y$-dependent terms in the KP equation. However, this is 
only partially true since the line solitons are known to be unstable 
against perturbations in the case of the KP-I equation. Since 
unavoidable numerical errors provide some form of perturbation, the 
propagation of (\ref{linesol}) also tests the $y$-dependent terms in 
the numerical implementation, see the example below. The reason for 
the use of a $y$-independent solution is that line solitons with an 
explicit $y$-dependence, \cf \cite{NMPZ} will not be periodic in $y$. 

\begin{figure}[!htb]
     \centering \includegraphics[width=10cm]{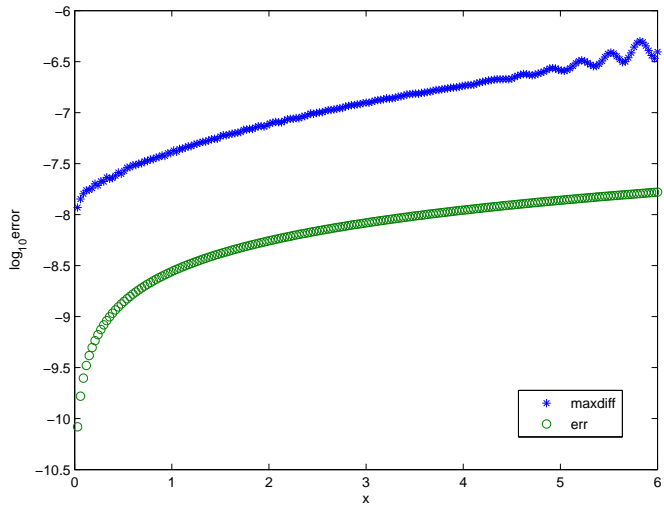}
    \caption{Numerical errors for the time evolution of 1-soliton 
    initial data: $L^{\infty}$ norm of the difference between 
    exact and numerical solution (maxdiff) and deviation from mass 
    conservation (err).}
    \label{figerror}
\end{figure}

The test of the propagation of the 1-soliton initial data is performed for the KP-I model in 
the following setting:
The $x$-coordinate takes values in the interval $[-\pi L_x,\pi L_{x}]$, where the length $L_{x}$ is 
always chosen such that the coefficients for the initial data are 
of the same order as the rounding error for high 
frequencies\footnote{MATLAB works internally
with a precision of $10^{-16}$. Due to rounding errors machine 
precision is generally limited to the order of $10^{-14}$.}. This 
reduces the error due to the non-periodicity of the initial data at 
the interval junctions to the order of the rounding error. The choice 
of the corresponding length $L_{y}$ is not important in this context, 
thus we shall choose it to be of the same value as $L_{x}$. We use 
the 1-soliton solution at time $t=0$ with $L_{x}=10$ and 
$x_{0}=-L_{x}$ in the initial data and determine for each time step the 
difference between the exact and the numerical solution. The computation 
is carried out with $N_{x}=2^{11}$ and $N_{y}=2^{7}$
modes (we will always use powers of 2 here since the FFT algorithm is 
most efficient in this case, but this is not necessary) and $4\times 10^3$ time 
steps for $t\in [0,6]$. The $L^{\infty}$ norm of the difference 
between the numerical and the exact solution is shown as a function 
of time in Fig.~\ref{figerror}. 

An alternative test of the numerical precision is provided by 
conserved quantities. Due to unavoidable 
numerical errors such quantities will be numerically a function of 
time. Typically the energy \eqref{energy} is a convenient quantity in this context. However the term 
$\partial_x^{-1} \partial_y u$ in general will have the same 
analytical properties as KP solutions not subject to the constraint (\ref{inconst}), 
namely a cusp along the $x$-axis. Since we use spectral methods, this non-analyticity of the integrand 
in (\ref{energy}) would lead to numerical errors, which however are mainly due to the 
way the integral in (\ref{energy}) is evaluated, and not to the  
numerical error in $u(t)$ itself. 

Therefore we shall instead consider the mass \eqref{mass}, which is 
also conserved in time, 
but which admits an integrand with better regularity properties than the energy.
Since mass conservation is not implemented, it provides a strong test for the accuracy of the code. We 
define the corresponding numerical error function by
\begin{equation}
    \mbox{err}(t):=1-\frac{{\mathrm M}[u(t)]}{{\mathrm M}[u_{\rm I}]}
    \label{errdef},
\end{equation}
where $\mathrm M[u(t)]$ is the numerically calculated mass 
which is obtained via FFT (in both spatial coordinates) of the integrand in (\ref{mass}) at each 
time step. For the example of the 1-soliton, this function 
is shown in Fig.~\ref{figerror}. It can be seen that the error 
obtained via the integral quantity is typically one order of magnitude 
higher than the maximal local difference of the exact and the numerical 
solution. In cases where no exact solution is known, we shall use 
mass conservation as an indicator of the precision of the numerics. 
Consequently, the number of modes and the time steps will always be chosen in a 
way such that the value of the function $\mbox{err}(t)$ is at least one order 
of magnitude lower than the precision of the numerical solution we 
are aiming at. 

\begin{figure}[!htb]
     \centering \includegraphics[width=14cm]{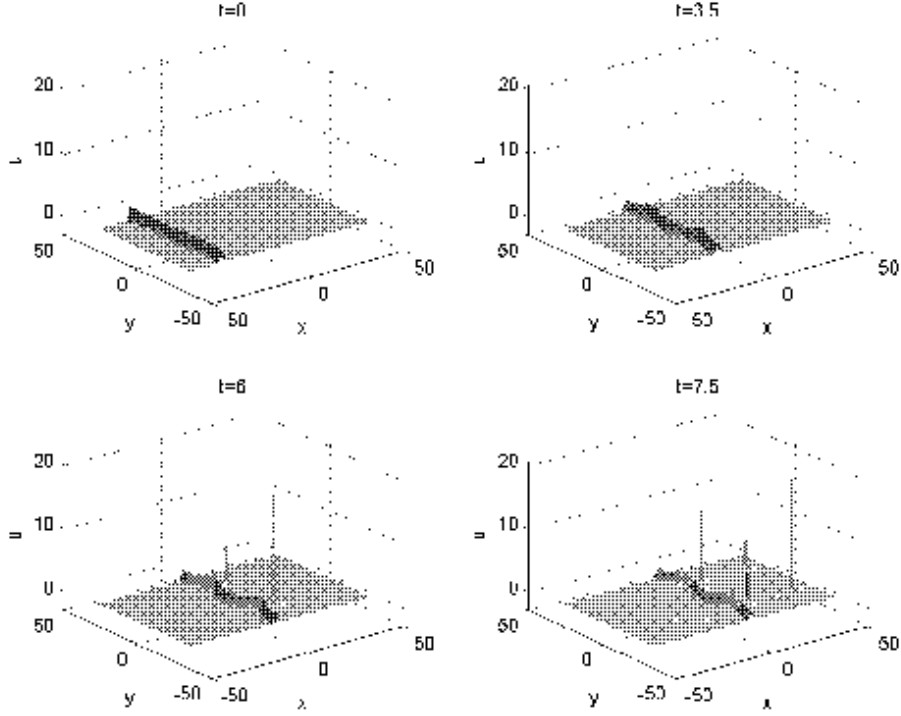}
    \caption{Time evolution of a perturbed line-soliton, growth of 
    the perturbation and formation 
    of lump solitons in the KP-I equation.}
    \label{figsolinst}
\end{figure}

As an example for the instability of the above given line-soliton in the case of the KP-I equation, 
we consider, as in \cite{SeIn}, a strongly 
perturbed line-soliton of \eqref{KP} with initial data 
\begin{equation}
    u_{\rm I}(x, y)= \frac{12} {\cosh^{2}(x-x_{0}+
    \delta \cos(0.2y))}.
    \label{initialsol}
\end{equation}
In Fig.~\ref{figsolinst} we show the growth of 
the perturbation with time and the formation of lump solitons for 
$\delta=0.4$, $L_{x}=12$, $L_{y}=10$, $\Delta t=1.33\times10^{-4}$ and $N_{x}=N_{y}=512$. Due 
to the algebraic decay of the lump solitons for $|x|,|y|\to\infty$, 
relative mass conservation is of the order of $10^{-3.5}$ in this 
case. This is the precision with which the code propagates lump 
solitons for the given parameters.

\begin{remark}
Notice that the formation of power-law tails in the time evolution of 
initial data with compact support will lead to a Gibbs phenomenon 
at the boundaries in our spatially periodic setting. These effects are 
small for sufficiently large periods, but clearly noticeable. Thus in 
the examples studied here, relative mass conservation will be of the 
order of $10^{-7}$, even for very small times, whereas in the corresponding 
computations for the KdV equation it is of the order of machine 
precision \cite{GrKl}. 
On the other hand, on the longer time scales considered here (when studying the 
rapid modulated oscillations we are interested in), it is still possible to achieve relative mass 
conservations of the order of $10^{-4}$. 
The same holds for the propagation of lump solitons 
such as \eqref{lump}, where the slow (algebraic) decay as $|x|,|y|\to \infty$ again leads 
to Gibbs phenomena at the boundaries. Nonetheless our scheme is able 
to propagate lump solitons on standard computers with relative mass 
conservations of the order of $10^{-4}$ as the code in \cite{FKM}. 
\end{remark}


\section{Modulation theory for small amplitude solutions}\label{smod}

Since the small dispersion limit $\epsilon \to 0$ of \eqref{KP} is 
only poorly understood, we first consider the analogous problem for solutions 
which are \emph{asymptotically small}, \ie of order $\O(\epsilon)$. 
To this end we perform a (formal) \emph{multiple scales expansion}, similar to those given in \cite{SuSu, ZaKu}, 
for such kind of solutions. 
Since the quadratic nonlinearity in this case is of lower order, we expect that the 
appearing fast oscillations can be described by a rather simple ansatz. 
More precisely they will be \emph{purely $x$-dependent}. Namely, via the 
so-called \emph{Airy equation} with plane wave initial data, \ie 
\be \label{airyeq}
\d_t v + \d_{xxx} v = 0, \quad v\big |_{t=0}= \E^{\I \eta x/\e}.
\ee
Note that this yields an asymptotic description which does \emph{not} involve fast 
oscillations in the $y$-variable, \cf below. This is certainly a simplification which nevertheless is justified by the 
computations given in Section \ref{slin} for the linear equation and by our numerical results. 

\subsection{Multiple scales expansion} 
To be more precise, let us consider the scaled KP equation 
\eqref{KP}, assuming that its solutions admit an asymptotic expansion of the form
\begin{align}\label{exp}
u(t,x,y) \sim \epsilon \, u_0 (t,x,y) + \epsilon^2\, u_1(t,x,y) + \O(\epsilon^3),\quad \e \ll 1.
\end{align}
As usual in multiple scales expansion methods, we consider 
\be
u_j=u_j(\tau, t,x,y, X,Y,T ), \quad \forall \, j\in \N,
\ee
introducing again the fast 
variables $X = x/\e$, $Y=y/\e$, $T =t/\e$, as well as the additional time-scale $\tau =\e t$. Moreover we 
assume $u_j$ to be \emph{periodic} w.r.t to $X,Y,T$. Note however that in the initial data \eqref{airyeq} we do not 
take into account the fast scale $Y=y/\e$. Thus we shall neglect this dependence in the following and 
one easily checks that this is consistent with our asymptotic expansion up to any order in $\e$.
Also note that we have to take into account additionally the (slowly varying) time scale $\tau =\e t$ 
in order to balance the nonlinearity with the term $\partial_x \partial_t u$.

Equating powers of $\e$ in our expansion, we find
that the leading order term $u_0$ is given by  
\be
\label{u0}
u_0(t,x,y)= \psi_0(\e t, x + 3 \eta^2 t,y) \, \E^{\I (\eta x+ \omega(\eta) t)/\e} + \mbox{c.c.}, 
\quad \omega(\eta) = \eta^3, 
\ee
for any given $\eta \in \R$, $\eta\not =0$, where $\psi_{0}(t,\xi(t,x),y)\in \C$ and ``c.c.'' refers to the complex conjugate. 
The fast oscillations are henceforth described by plane waves propagating according to 
$\omega(\eta)= \eta^3$, the dispersion relation of the Airy equation. 
Moreover, in \eqref{u0}, we need to take into account a \emph{drift} defined by 
$$ 
{\xi}(t,x) :=  x + 3  \eta^2  t,
$$ 
\ie we need to perform a change of variables into a reference frame moving at group velocity $v_{\rm g} = 3 \eta^2$. 
Thus, on the so-called \emph{ballistic scales} $(t, x)=(\e T, \e X)$, the slowly varying amplitude satisfies the 
transport equation
\be
\label{trans}
\partial_t \psi_0 - 3\eta^2  \, \partial_{x} \psi_0 =0,
\ee
which is obtained from the solvability condition for terms of order $\O(\e)$. 

Consequently one finds that the fast scale dynamics in the first order corrector $u_1$ is determined via
\be
\partial_X {\left(\partial_T  + \partial_{XXX}  \right)}\, u_1 = -\frac{1}{2} \, 
\partial_{XX}\left(\psi^2_0(\tau,\xi(t,x),y) \E^{ 2\I(\eta X  + \omega(\eta) T)} + \mbox{c.c.}\right)
\ee
and thus $u_1$ is given by
\be
\label{u1}
\begin{split}
u_1(t,x,y)= & \ \frac{1}{6\eta^2}\, \psi^2_0(\tau,\xi(t,x),y) \, \E^{ 2\I(\eta X  + \omega(\eta) T)} 
+ \psi_1(\tau,\xi(t,x),y) \, \E^{\I(\eta X  + \omega(\eta)T)} \\
& \ + \mbox{c.c.} + \phi(\tau,\xi(t,x),y){\Big |}_{\tau=\e t, X= x/\e, T=t/\e},
\end{split}
\ee
where $\phi$ can now be interpreted as a \emph{mean field} generated by 
$\psi_0$. From the mathematical point of view it is necessary to take into account this 
non-oscillatory field to balance the terms generated via the quadratic nonlinearity. 
We eventually find that $\psi_0$ and $\phi$ 
solve a coupled system of \emph{Davey-Stewartson type} (DS). 
More precisely we get
\be \label{DS}
\left \{
\begin{aligned}
\I \, \d_\tau \psi_0 - 3\eta \, \d_{\xi \xi} \psi_0 + \frac{\lambda}{\eta} \, \d_{yy} \psi_0
- \left(\frac{1}{6\eta}\, |\psi_0|^2 + \eta \phi \right)\psi_0 = 0, \\
3 \eta^2 \, \d_{\xi \xi} \phi + \lambda \, \d_{yy} \phi + \d_{\xi \xi}|\psi_0|^2 = 0.
\end{aligned}
\right.
\ee
This mean field system describes the dispersive effects which are visible on 
time-scales of order $\O(\epsilon t)$. 
Analogous to the KP equation, the DS system represents a completely integrable model in $d=2$ spatial dimensions. 
The case where, for $\eta >0$, $\lambda = + 1$, is referred to as the ``hyperbolic-elliptic'' case, \cf \cite{SuSu}, 
whereas for $\lambda = -1$ we are in the so-called ``elliptic-hyperbolic'' situation. 
Here we shall only focus on the former 
case, induced by the KP-II model, as the latter requires a special numerical treatment due to the 
appearing wave-type operator in the second equation of \eqref{DS}. 
Thus we require, that the mean field behaves like  
\be
\phi(\tau, \xi, y) \to 0, \quad \mbox{as $|\xi|,|y| \to \infty$.}
\ee
\begin{remark} 
An analogous asymptotic expansion for the KdV equation 
yields a cubic nonlinear Schr\"odinger equation, instead of the DS system, as has been rigorously 
proved in \cite{Sch}. We expect that a similar analytical strategy could also be applied in our case 
to rigorously establish the above given formal asymptotics.  
\end{remark}

In the present work, the numerical algorithm used to solve \eqref{DS} is analogous to the one used 
for the KP equation (for an alternative approach see \cite{BMS}). 
More precisely, denote by $\widehat{\psi_0}(\tau, k_\xi,k_y)$ the Fourier 
transform of $\psi_0(\tau,\xi,y)$ and let $\widehat{\psi_0|\psi_0^{2}|}$ 
and $\widehat{\psi_0\phi}$ be the Fourier 
transforms of $\psi_0|\psi_0^{2}|$ and $\psi_0\phi$, respectively. 
Thus we get for the system (\ref{DS})
\begin{equation}
    \left \{
\begin{aligned}
    \mathrm{i}\partial_{\tau}\widehat{\psi_0}+\left(3\eta k_\xi^{2}-
    \frac{\lambda}{\eta}k_{y}^{2} \right) \widehat{\psi_0} 
    -\frac{1}{6\eta}\, \widehat{\psi_0|\psi_0^{2}|}-\eta \widehat{\psi_0\phi}
    =0,\\
    \widehat{\phi}= -\frac{\eta k_\xi^{2}}{3\eta^{2}k_\xi^{2}
    +\lambda k_{y}^{2}} \, \widehat{|\psi_0^{2}|},
\end{aligned}    
\right.
    \label{DSFT}
\end{equation}
with $\eta >0$. Again we shall use an integrating factor to avoid a stiff part in the
first equation of (\ref{DSFT}), \ie 
\begin{equation}    
    \partial_{\tau} \left(\mathrm{e}^{-\mathrm{i}\tau(3\eta k_\xi^{2}-
    \lambda k_{y}^{2}/\eta)}\widehat{\psi_0} \right) +\mathrm{i}
    \mathrm{e}^{-\mathrm{i}\tau(3\eta k_\xi^{2}-
	\lambda k_{y}^{2}/\eta)}\left(\frac{1}{6\eta}\, \widehat{\psi_0|\psi_0^{2}|}
	+\eta \widehat{\psi_0\phi}\right)
    =0
\label{DSint}.
\end{equation}
The time integration will be carried out as before with a fourth order 
Runge-Kutta method. Accuracy of the code will be checked by 
conservation of the \emph{wave energy}
\be\label{wenergy}
\mathrm N[\psi_0(\tau)]:=\int_{\R^2} | \psi_0(\tau ,\xi ,y)|^2 \, \D \xi \, \D y , 
\ee
which is preserved in time for the DS system \eqref{DS}.  

\subsection{Numerical examples}
To study a concrete example we consider, for $\eta = 1$, real-valued initial data to the DS system \eqref{DS} in the following form 
\begin{equation}
    \psi_0\big|_{t=0}\equiv \psi_{\rm I}(x,y) 
    =-\partial_{x}\, \mbox{sech}^{2}(R),\quad R:= \sqrt{x^{2} +y^{2}}
    \label{ds1}.
\end{equation}
The choice for these initial data is motivated by the fact that they are smooth, localized in $x,y$, 
and that they could, in principle, also serve as initial data for the KP equation \eqref{KP}, since they 
satisfy the constraint (\ref{inconst}).  

\begin{figure}[!htb]
      \centering \includegraphics[width=14cm]{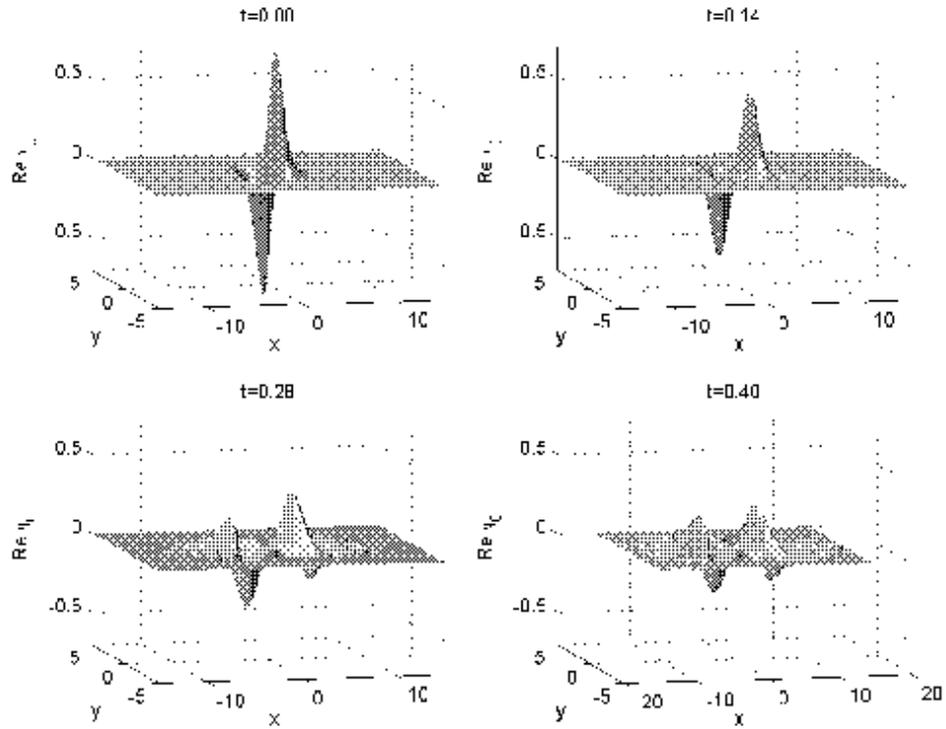}
    \caption{Time evolution of $\re \psi_{0}(t)$, solution to the 
    DS system 
    with initial data (\ref{ds1}), shown for several values 
    of $t\geq0.$}
    \label{figdst}
\end{figure}

\begin{figure}[!htb]
      \centering \includegraphics[width=10cm]{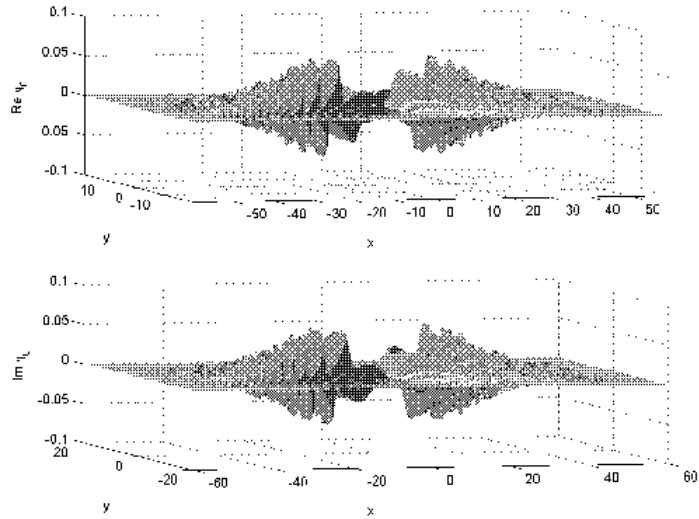}
    \caption{$\re \psi_0(t)$ and $\im \psi_0(t)$, obtained from the initial data (\ref{ds1}), at time $t=2$.}
    \label{figdst2}
\end{figure}

\begin{figure}[!htb]
      \centering \includegraphics[width=10cm]{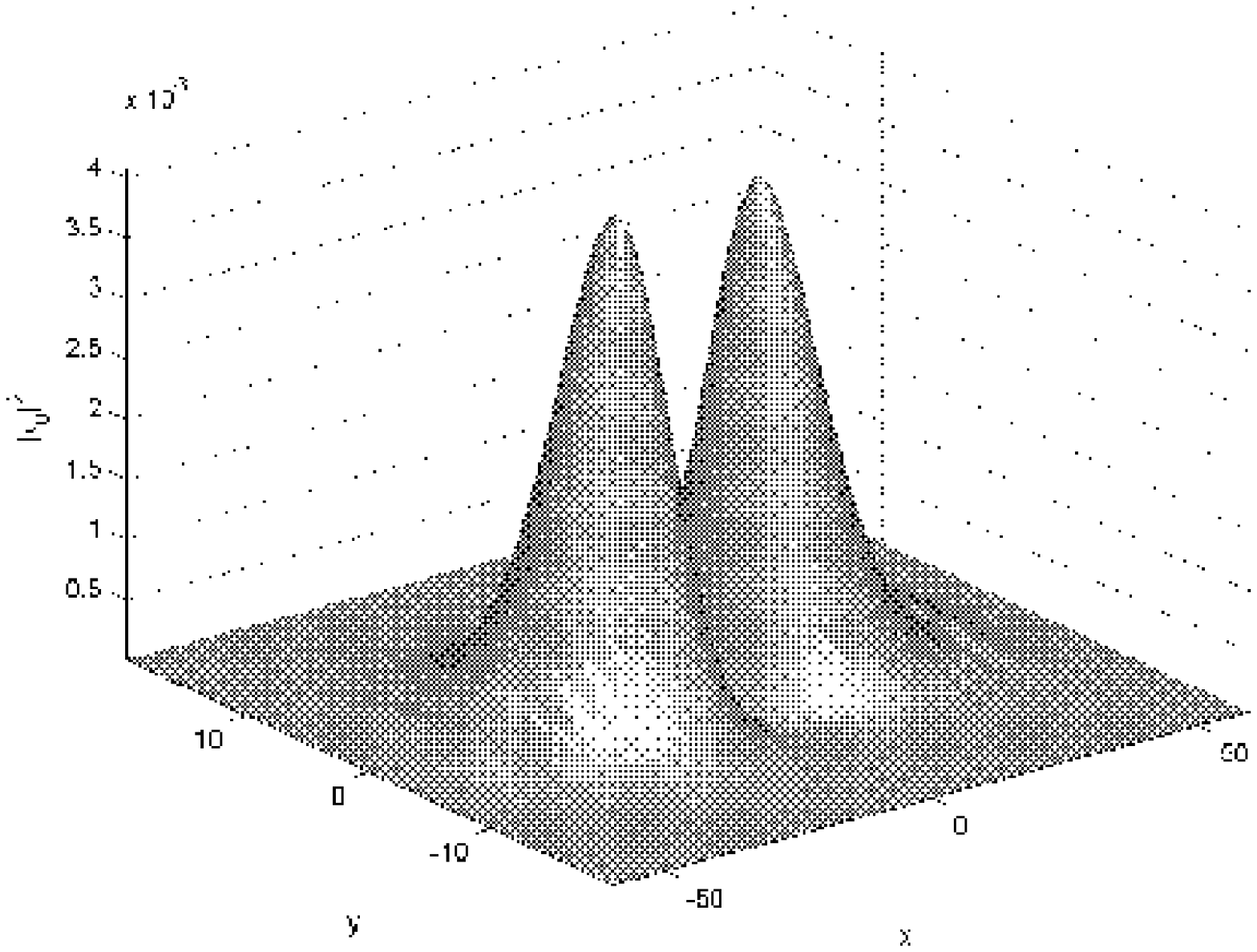}
    \caption{$|\psi_0(t)|^{2}$, 
    obtained from the initial data (\ref{ds1}), at time $t=2$.}
    \label{figdst2abs}
\end{figure}

We then solve the DS system (\ref{DS}) with initial data (\ref{ds1}) for 
$t\in [0,0.4]$ with $N_{x}=N_{y}=512$ and $\Delta t=2\times10^{-3}$. 
The wave energy is conserved in this computation up to 
the order of machine precision (the error is smaller than 
$10^{-13}$). We show the real part of the function $\psi_{0}(t)$ for 
several values of $t$ in Fig.~\ref{figdst}.
It can be seen that the initially localized pulse spreads and exhibits oscillations, 
both mainly in $x$-direction. For longer times, the modulation of the 
DS solution becomes more pronounced as can be seen in 
Fig.~\ref{figdst2}.
Notice that the real and imaginary part of $\psi_0(t)$ have almost the same envelope, 
they mainly differ by a phase shift. This is more obvious by 
considering the absolute value of $\psi_{0}$ in Fig.~\ref{figdst2abs} 
which virtually shows no modulations.

Corresponding to \eqref{ds1}, the KP-II equation is then solved with initial data 
\begin{equation}
    u_{\rm I}(x,y)=2\epsilon\, \psi_{\rm I}(R) \cos\left(\frac{x}{\epsilon}\right),\quad R=\sqrt{x^{2} +y^{2}}.
    \label{kpdsinitial}
\end{equation}
Note however, that these initial data do \emph{not} satisfy the constraint (\ref{inconst}). 
To enforce \eqref{inconst} we numerically 
compute the Fourier transform of $u_{\rm I}$ and set the Fourier 
coefficients of all terms corresponding to $k_{x}=0$ in these data equal to zero. 
This can be justified by the fact that in our examples, where we take $\epsilon = 0.1$ and smaller, 
these Fourier coefficients are only of the order $10^{-14}$, \ie of the order of the rounding 
error. Thus, within numerical precision, we can directly use 
these initial data, which are now adapted to both, the asymptotic expansion and the integral constraint \eqref{inconst}. 
For the precise parameters used in the computations of the KP equation, for times between 0 and 4, see 
table \ref{tab1}. 
\begin{table}[tbp]
    \centering
    \begin{tabular}{|c|c|c|c|c|c|c|}
	\hline
	$-\log_{10}\epsilon$ & $\log_{2}N_{x}$ & $\log_{2}N_{y}$ 
	&$\Delta t$ & $L_{x}$ & $L_{y}$ & $\log_{10}err$   \\
	\hline
	1 & 10 & 7 & $8\times10^{-5}$ & 10& 10& 4.98  \\
	\hline
	1.25 & 11 & 7 & $8\times10^{-5}$ & 10& 10& 4.23  \\
	\hline
	1.5 &  12 & 7 & $6.67\times10^{-5}$& 10& 10& 4.21  \\
	\hline
	1.75 & 12 & 7 & $8\times10^{-5}$ & 5& 5& 4.07  \\
	\hline
	2 & 12 & 7 & $6.67\times10^{-5}$ & 5& 5& 4.64  \\
       \hline
    \end{tabular} 
    \caption{Parameters in the computation of the KP solutions}
    \label{tab1}
\end{table}

\begin{figure}[!htb]
      \centering \includegraphics[width=14cm]{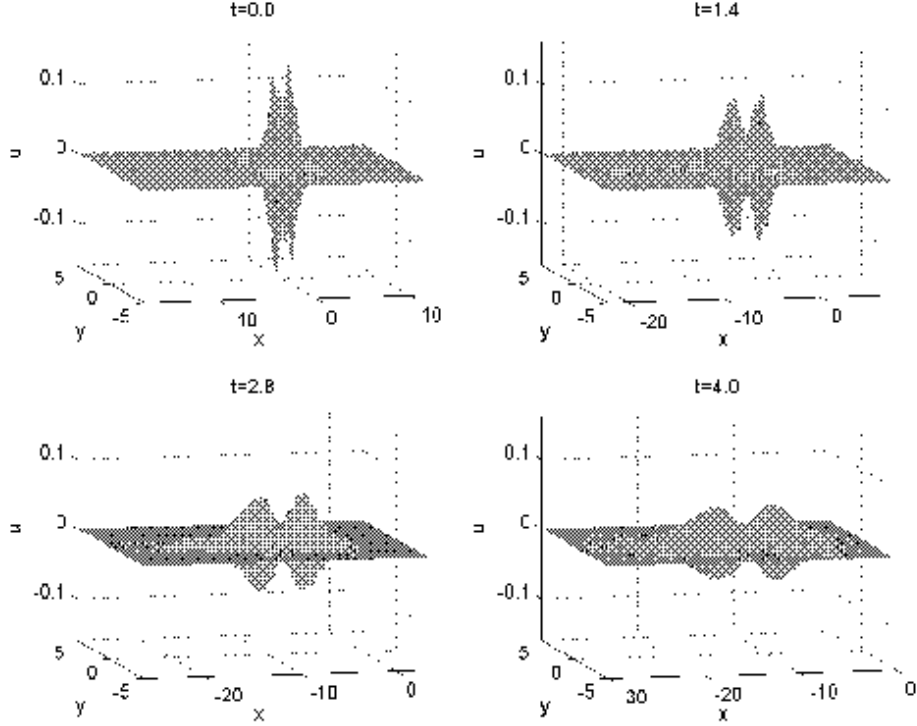}
    \caption{Time evolution of the initial data (\ref{kpdsinitial}), 
    governed by the KP-II equation 
    with $\epsilon=0.1$, shown for several values of $t\geq 0$ and shifted by $x\to x-12$.}
    \label{figkpdst}
\end{figure}

In Fig.~\ref{figkpdst} we plot the solution of the KP equation for $\epsilon=0.1$ for several 
values of $t$.
As expected the pulse travels roughly with the velocity 
$v_{\rm g}=3$ (notice that the shown region in the plots is co-moving as 
can be seen from the $x$-scales) and the initially localized pulse 
spreads mainly in $x$-direction. Since we only consider small 
amplitudes here, the characteristic KP tails are hardly visible on the studied 
timescales. Thus the periodic boundary conditions do not influence 
the model.

The above initial data are of the required form for the asymptotic expansion (\ref{exp}) 
in leading order of $\e$. In principle one could try to obtain a 
refined asymptotics, including higher order 
corrector terms. This however yields the highly nontrivial problem of satisfying the 
KP constraint (\ref{inconst}) up to sufficient high order. More precisely if one takes into account 
also the first order corrector \eqref{u1}, then, since $\widehat{\phi}(t)$ 
is proportional to the Fourier transform of $|\psi_{0}(t)|^{2}$, the 
coefficients for $k_{x}=0$ are of the order $\O(\epsilon^{2})$. In order to 
satisfy the constraint also up to $\O(\epsilon^{2})$ errors, it consequently would 
be necessary either to choose $\psi_{2}(0,x,y)$ in an appropriate way, or to 
consider a function $\psi_{\rm I}(x,y)$ which is subject to a nonlinear 
integral constraint following from (\ref{inconst}). 
In the following we shall simply neglect higher order corrector terms in the initial data as 
we only aim (numerically) for the leading order asymptotic description. 
 
\begin{figure}[!htb]
      \centering \includegraphics[width=14cm]{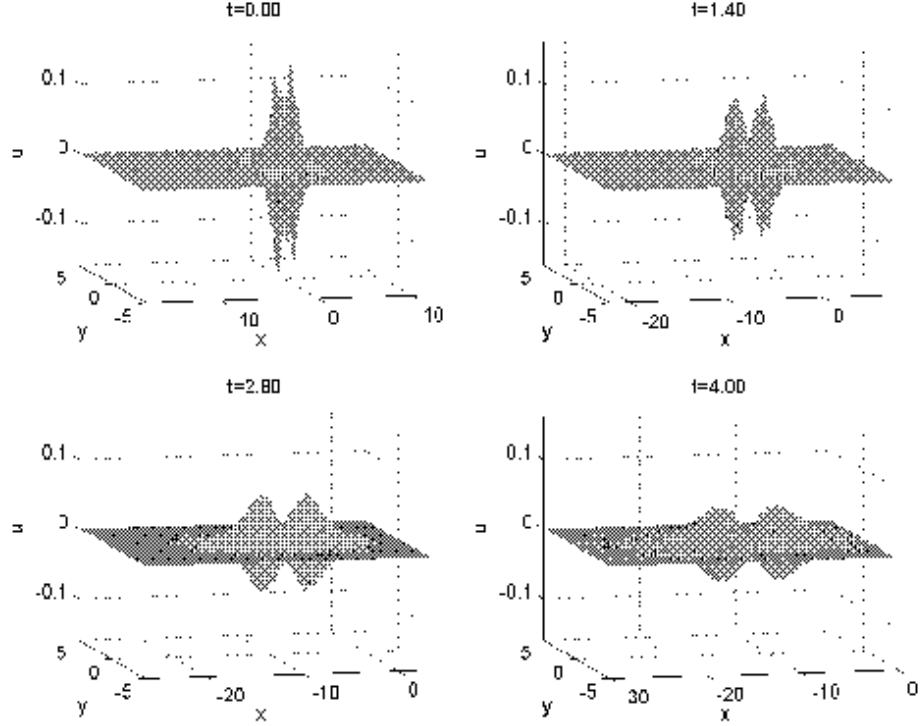}
    \caption{The asymptotic solution $u_{\rm app}(t,x,y)=2\epsilon \, \re \left(\psi(\epsilon t,\xi(t,x),y)\E^{\I(x+t)/\epsilon}\right)$ 
    approximating the true KP-II solution of Fig.~\ref{figkpdst}, shown for several values of $t\geq0$.}
    \label{figdskpt}
\end{figure}

\begin{figure}[!htb]
      \centering \includegraphics[width=10cm]{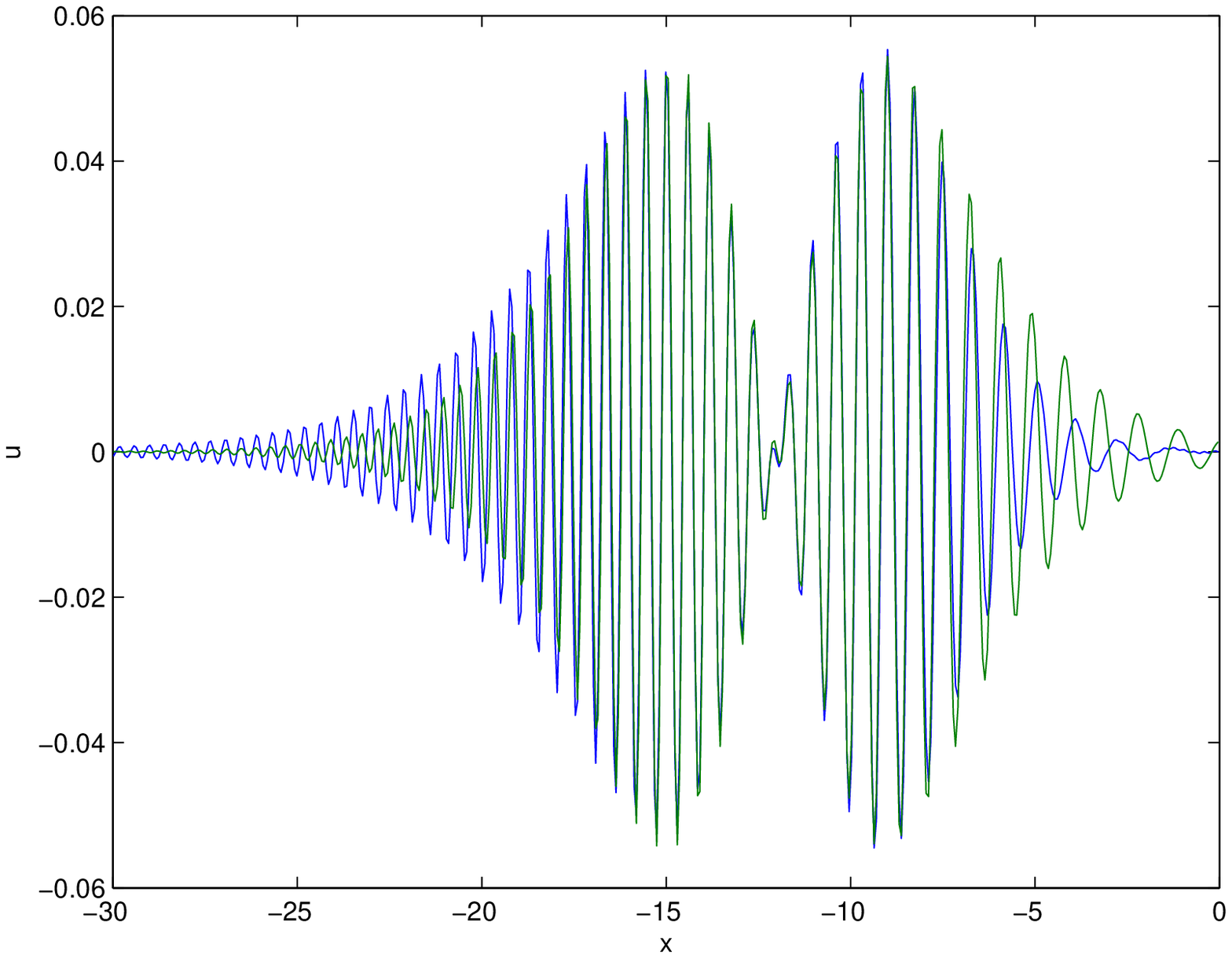}
    \caption{Solution to the KP-II equation with initial data 
    (\ref{kpdsinitial}) (blue) and the corresponding asymptotic 
    solution (green) for $y=0$, $t=4$ , and $\epsilon=0.1$.}
    \label{fig2in1kpds}
\end{figure}

\begin{figure}[!htb]
      \centering \includegraphics[width=10cm]{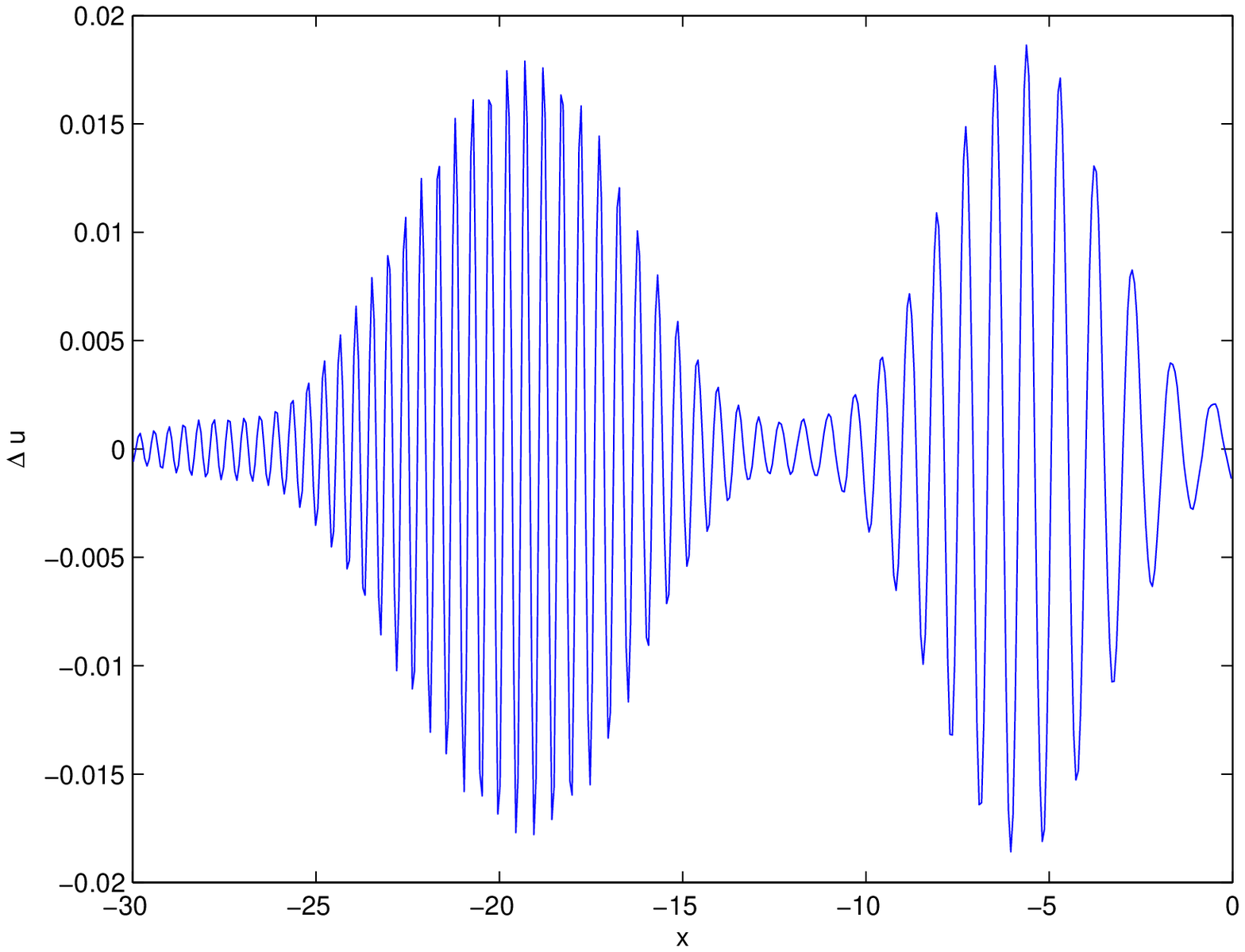}
    \caption{Difference of the solution to the KP-II equation 
    with initial data 
    (\ref{kpdsinitial}) and the corresponding asymptotic 
    solution for $y=0$, $t=4$, and $\e = 0.1$.}
    \label{figdeltakpdsy}
\end{figure}

\begin{figure}[!htb]
      \centering \includegraphics[width=10cm]{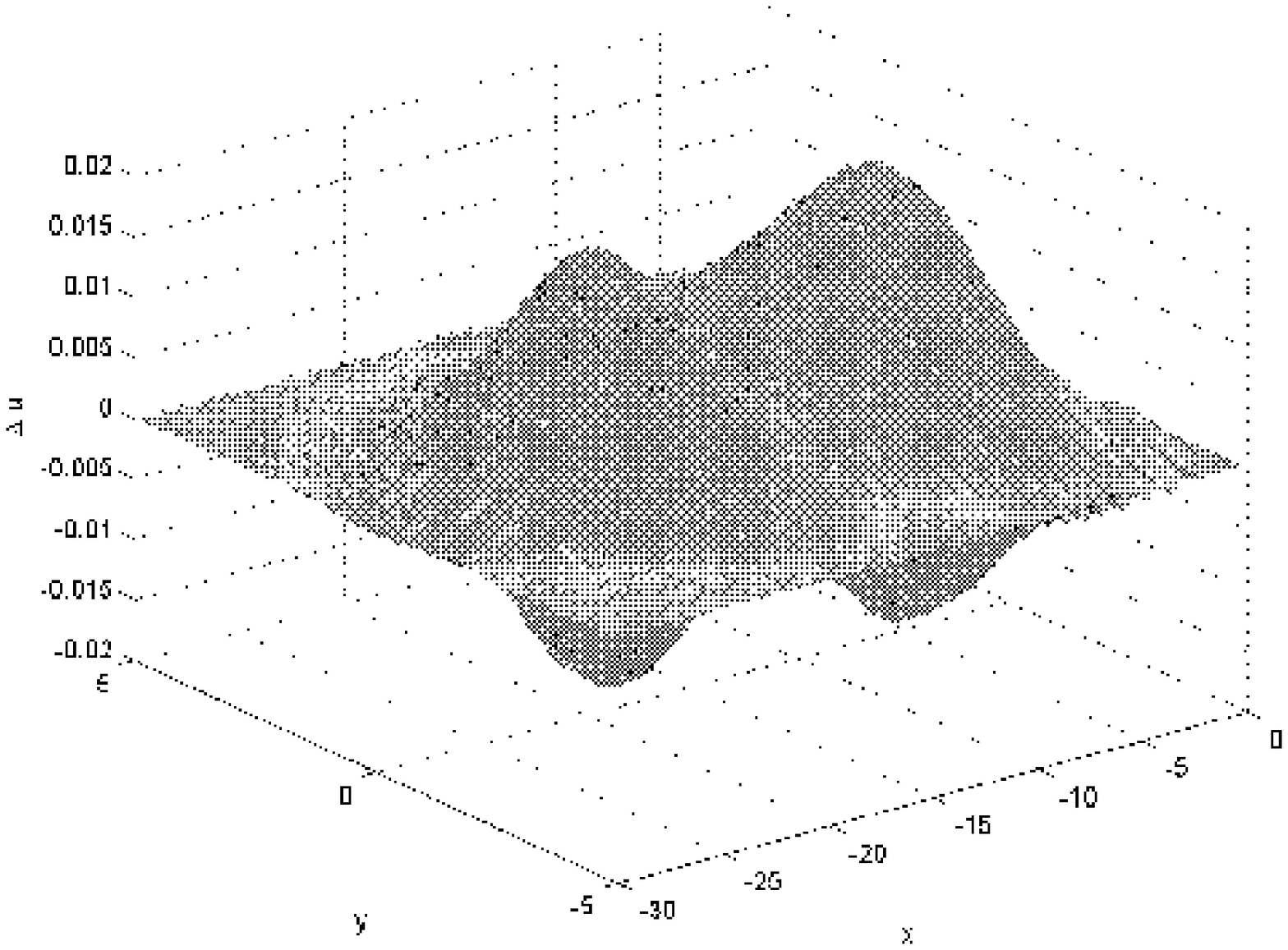}
    \caption{Difference of the solution to the KP-II equation 
    with initial data 
    (\ref{kpdsinitial}) and the corresponding asymptotic 
    solution for $t=4$ and $\epsilon=0.1$.}
    \label{figdeltakpds}
\end{figure}

\begin{figure}[!htb]
      \centering \includegraphics[width=10cm]{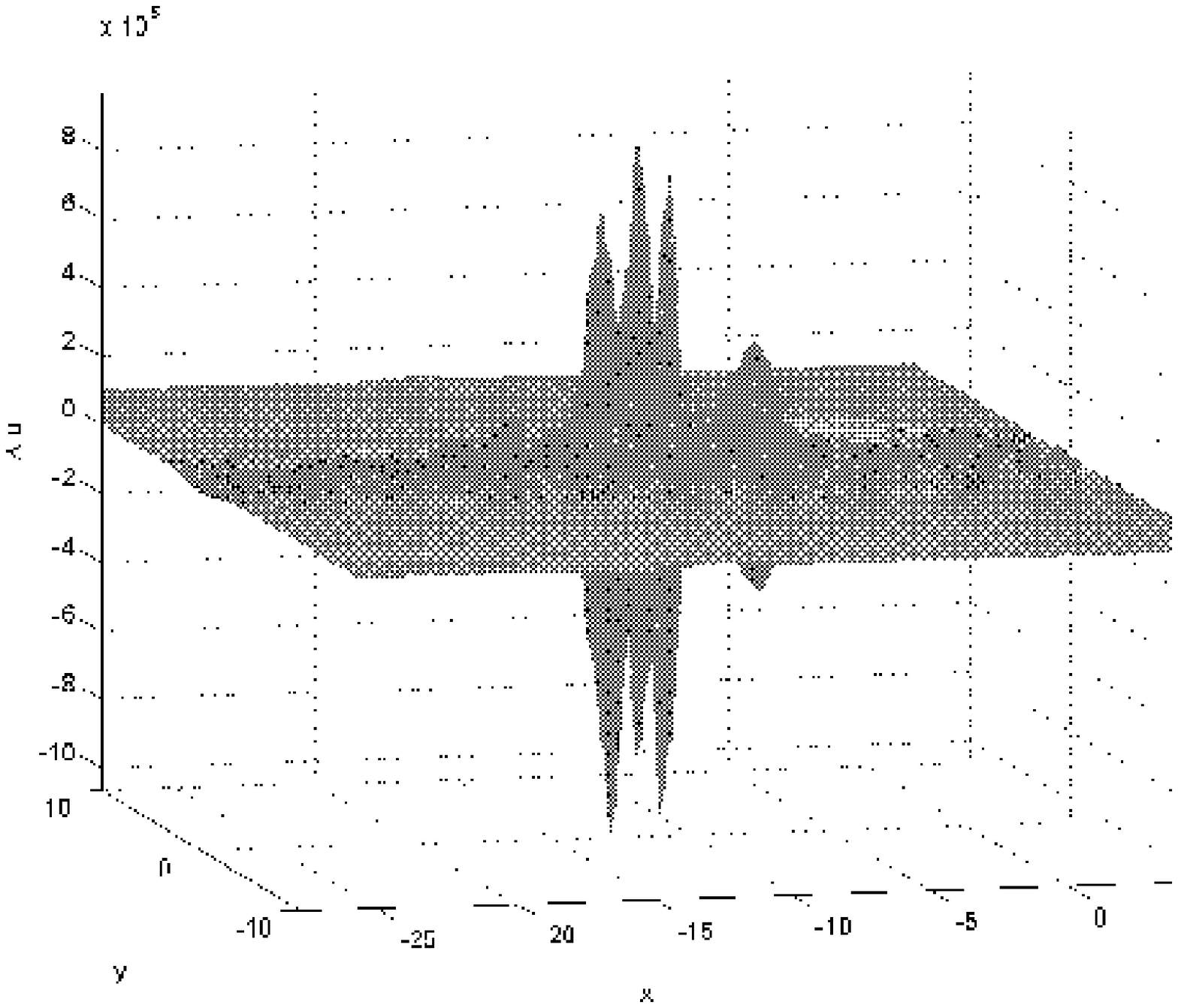}
    \caption{Difference of the solution to the KP-II equation 
    with initial data 
    (\ref{kpdsinitial}) and the corresponding asymptotic 
    solution for $t=4$, and $\epsilon=10^{-2}$.}
    \label{figdeltau1e4}
\end{figure}

\begin{figure}[!htb]
      \centering \includegraphics[width=10cm]{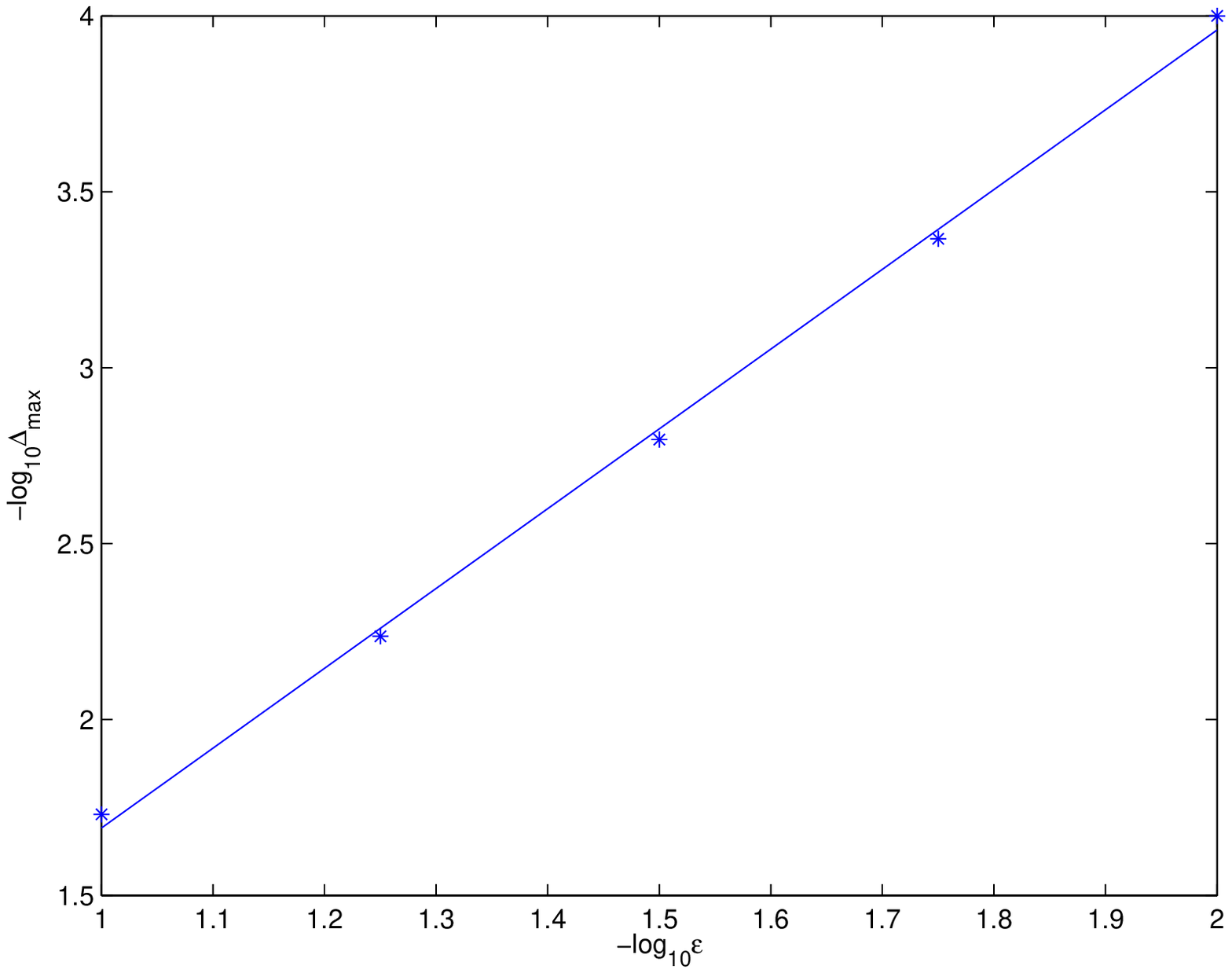}
    \caption{Error $\Delta_{\infty}$ for several 
    values of $\epsilon$. The data can be fitted by least square 
    analysis with a straight line 
    $-\log_{10}\Delta_\infty=-a\log_{10}\epsilon+b $ with $a=2.27$ and 
    $b=-0.58$. 
The correlation coefficient is $r=0.999$, the standard error for $a$ 
is $\sigma_{a}=0.08$.}
    \label{figdeltamax}
\end{figure}

The above given expansion indicates that the solution to the DS 
system \eqref{DS}, with initial data $\psi_{\rm I}$, should approximate, via (\ref{exp}) and 
(\ref{u0}), the corresponding KP solution up to errors of order 
$\mathcal{O}(\epsilon^\alpha)$, for some $\alpha >1$, on any finite time-interval. 
In Fig.~\ref{figdskpt}, we plot the approximating solution 
\be \label{approx}
u_{\rm app}(t,x,y)=2\epsilon \re \, \left(\psi(\epsilon t,\xi(t,x),y)\E^{\I(x+t)/\epsilon}\right),
\ee
for several values of $t$.
It can be seen that this asymptotic solution gives the expected 
description of the KP solution, which is even better visible in 
Fig.~\ref{fig2in1kpds}, where we have plotted the two solutions for 
$t=4$ and $y=0$ in one frame.
The difference of these two solutions for $t=4$ can be seen for $y=0$ in 
Fig.~\ref{figdeltakpdsy} and in the whole $(x,y)$-plane in
Fig.~\ref{figdeltakpds}.
\begin{remark} Notice the 
difference between the KP solution and the asymptotic solution for 
different signs of the co-moving coordinate $\xi(t,x)$. We chose initial 
data which are odd in $x$. Obviously the KP equation 
does not conserve the parity of the initial data, whereas the DS 
system does. In our example the asymptotic solution is, up to the leading order 
in $\epsilon$, an odd function in $x$, whereas this is no longer true for the KP solution. 
Higher order terms in the asymptotic solution, which we neglected here, will 
break this symmetry. 
\end{remark}
For smaller values of $\epsilon$ the approximation gets better, but the numerical resolution 
of the high frequencies becomes increasingly more difficult. 
The main problem is 
that the DS system approximates the KP solution only on the time scale $\epsilon t$. 
Thus in order to see the effects of the modulations due to the DS system within the KP solution 
one would need to solve the KP equation for extremely long times if $\e$ is chosen very small. This 
is of course numerically rather expensive, in particular in our $2+1$ 
dimensional case. We will therefore limit our analysis to the range 
$0.01\leq \epsilon \leq0.1$. The above given example however clearly 
illustrates the applicability of the 
asymptotic expansion. For $\epsilon=0.01$ the point wise difference between the 
KP solution and the asymptotic solution is, as expected, at most of 
the order $10^{-4}$ and thus barely one order of magnitude higher than 
the numerical error, \cf Fig.~\ref{figdeltau1e4}.
To get more insight we plot the $L^{\infty}(\R^2)$ norm of the difference, 
denoted by $\Delta_{\infty}(t)$, in dependence of $\epsilon$
in Fig.~\ref{figdeltamax}. This error is found to be \emph{monotonically} increasing in time and furthermore it 
decreases roughly like $\epsilon^{9/4}$. Indeed the data obtained at the last time-step $t=4$, 
can be fitted by a straight line 
$-\log_{10}\Delta_{\infty}=-a\log_{10}\epsilon+b $ with $a=2.27$ and 
$b=-0.58$. 
The correlation coefficient is then found to be $r=0.999$, the standard error for $a$ 
is $\sigma_{a}=0.08$. 

\begin{figure}[!htb]
      \centering \includegraphics[width=10cm]{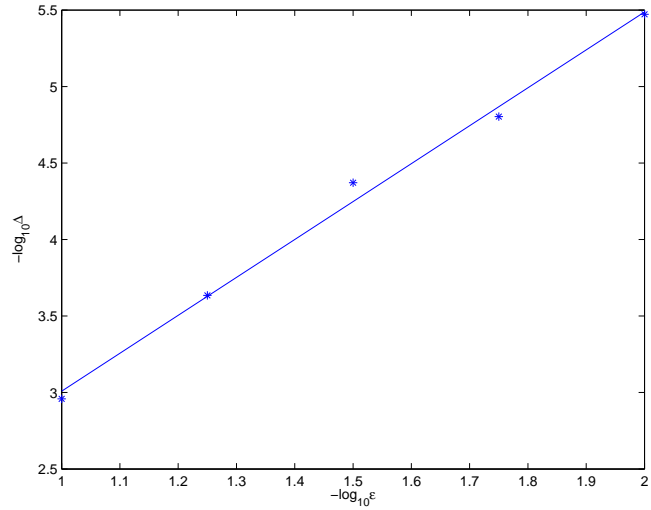}
    \caption{Error $\Delta_2$ as defined in \ref{delta} for several 
    values of $\epsilon$. The data can be fitted by least square 
    analysis with a straight line 
    $-\log_{10}\Delta_2=-a\log_{10}\epsilon+b $ with $a=2.48$ and 
    $b=0.53$. 
The correlation coefficient is $r=0.997$, the standard error for $a$ 
is $\sigma_{a}=0.17$.}
    \label{figloglog}
\end{figure}

To define an integral quantity $\Delta_2(t)$ for the error between the true KP solution and its (leading order) 
asymptotic description \eqref{approx}, we also consider the $L^{2}(\R^2)$ 
norm of the difference. Indeed this is the most widely used definition of an asymptotic error in such singular limiting regimes, 
\cf \cite{Sch} for the analogous definition in the KdV case. 
Thus we integrate the square of the difference of these solutions over the 
domain of definition via an FFT (in both variables) and normalize this quantity 
dividing it by the area of the fundamental domain, \ie 
\begin{equation}
    \Delta_2(t): = \frac{1}{2\pi \sqrt{L_{x}L_{y}}} \left(\int_{-\pi L_{x}}^{\pi L_{x}}\int_{-\pi 
    L_{y}}^{\pi L_{y}}\left(u(t,x,y) -u_{\rm app}(t,x,y)\right)^{2} \D x\,\D y\right)^{1/2}
    \label{delta}.
\end{equation}
We find that $\Delta_2(t)$ again increases monotonically in time.
For $t=4$ we get the values of $\Delta_2$ shown in Fig.~\ref{figloglog}. By 
linear regression the data can again be fitted by a straight line 
$-\log_{10}\Delta_2=-a\log_{10}\epsilon+b $, with $a=2.48$ and $b=0.53$. 
The correlation coefficient in this case is $r=0.997$ and the standard error for $a$ 
is $\sigma_{a}=0.17$. 
\begin{remark} 
It is not surprising that $\Delta_2$ decreases faster with $\epsilon$ 
than $\Delta_{\infty}$ since not only do the amplitudes shrink 
with $\epsilon$, but also the pulse is more localized in $x$ and 
$y$. Thus a smaller integral error is to be expected. 
\end{remark}
In other words we get that \emph{numerically} the $L^2(\R^2)$ difference between the true 
KP solution and its asymptotic description is approximately of the \emph{order} $\O(\e^{5/2})$. 
This is remarkable since it fits with the analytical results of \cite{Sch} where the analogous limit from KdV to 
the cubic nonlinear Schr\"odinger equation is considered. 
\begin{remark}
Note however, that for a comparison of our asymptotic description \eqref{u0} 
with the one given in \cite{Sch} one has to take into account a rescaling of the spatial scales 
on which the modulation amplitudes varies. Thus rescaling $x$ and $y$ 
(the latter scale is of course not present in \cite{Sch}; 
one has to plug this in ``by hand'') such that the two descriptions match each other,
one realizes that an additional factor $\e$ has to be taken into account 
in order to conserve the $L^2(\R^2)$ norm of the solutions. 
In summary one checks that Theorem 1 in \cite{Sch} yields an asymptotic error of the order $\O(\e^{5/2})$, 
\ie \emph{exactly as in our case}.
\end{remark}

\section{Small dispersion KP solutions with amplitudes of order $\O(1)$}\label{slow}

\begin{figure}[!htb]
      \centering \includegraphics[width=10cm]{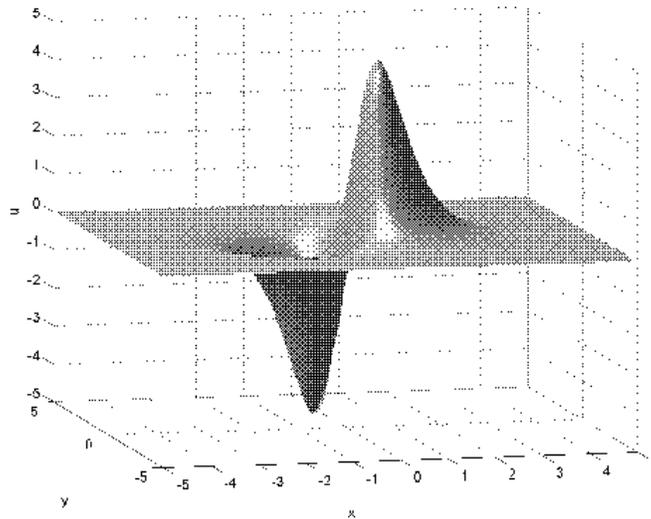}
    \caption{Initial conditions for the KP equation.}
    \label{initial}
\end{figure}

In this section we will study numerically solutions to the KP 
equation in the regime of small dispersion $\epsilon \ll 1$. In contrast to Section \ref{smod} 
though we shall not restrict ourselves to small amplitude solutions. Rather we shall consider smooth initial 
data with an amplitude of order $\O(1)$, as $\e \to 0$. To this end let us first discuss briefly the 
resulting (formal) limiting equation \eqref{dKP} and its numerical implementation in the next subsection.
We shall then investigate a concrete example with initial data of the form 
\begin{equation}
u_{\rm I}(x,y)=-6\, \partial_{x}\mbox{sech}^{2}(R),\quad R= \sqrt{x^{2}+y^{2}}
\label{kpld1},
\end{equation}
as can be seen in Fig.~\ref{initial}.
Note that these data satisfy the constraint (\ref{const}). 
The factor 6 in (\ref{kpld1}) is included in order to be able to compare our results with those given in \cite{GrKl} 
for the KdV equation, where a different definition of the amplitude $u(t)$ 
was used (the $u(t)$ there corresponds to $6u(t)$ here). With the introduction of the factor 6 we 
expect the same dynamical time scales in the KdV sector of \eqref{KP} as in \cite{GrKl}. 

\subsection{The dKP equation and its dissipative regularization}\label{subsdKP}

We do not expect the formal limiting dKP equation \eqref{dKP}, \ie 
$$
\partial_x {\left(\partial_t u + u \, \partial_x u \right)} + \lambda \, \partial_{yy} u =0,
$$
to be the correct description of the dispersionless limit for \eqref{KP}, at 
least not for all times $t\geq 0$. This believe stems from the closely related 
situation encountered in the 
\emph{inviscid Burgers}, or \emph{Hopf equation}, given by
\be \label{Bu}
\partial_t u + u \, \partial_x u  =0, \quad u\big |_{t=0} = u_{\rm I} (x) .
\ee
Equation \eqref{Bu} can be seen as the (formal) \emph{dispersionless limit} of the KdV equation. 
In general though this only holds for some finite time $0\leq t_c <\infty$. More precisely, the 
description of the disperionless KdV limit via \eqref{Bu} fails, after the appearance of the 
first shock in the solution of \eqref{Bu}. The 
corresponding \emph{break time} is given by  
\be
t_c = \min_{x_0 \in \R} \left(-\frac{1}{\partial_x u_{\rm I}(x_0)}\right),
\ee
which can be easily seen when solving \eqref{Bu} by the method of characteristics, yielding
\be \label{hopf}
u(t,x)=u_{\rm I} (x_0),\quad x=  u_{\rm I}(x_0) t+x_0,
\ee 
the so-called \emph{Hopf solution}. Consequently, we also expect in the case of the dKP equation that 
a shock will be formed after some finite time. (Clearly the behavior 
of the dKP equation 
is completely analogous to the Burger's case, if one considers $y$-independent solutions.) 
On the other hand, for small, but still finite, $\e\ll1$, this 
``dKP-shock'' presumably will be smoothed out by rapid oscillations, similar to the 
KdV case, \cf the numerical examples below. 
\begin{remark} 
To convince ourselves that, apart from the KdV sector, the concept of shocks is really 
applicable in the dKP equation we remark that its principal part is given by 
\be
Pu:=\partial_{xt} u + u \, \partial_{xx} u + \lambda \partial_{yy}u. 
\ee
As one easily checks, the coefficient matrix in this partial differential 
operator $P$ admits \emph{three, distinct, real eigenvalues} given by 
\be
\mu_{1,2}=\frac{1}{2}\left(u\pm \sqrt{u^2+1}\right),\ \mu_3 = \lambda.
\ee
Thus the dKP equation indeed falls into the class of second order hyperbolic PDEs.c
However the associated initial value problem as considered in this work, furnishes a characteristic Cauchy problem, 
a fact which has already been discussed in Subsection \ref{sub}.
\end{remark}
 
\begin{figure}[!htb]
      \centering \includegraphics[width=10cm]{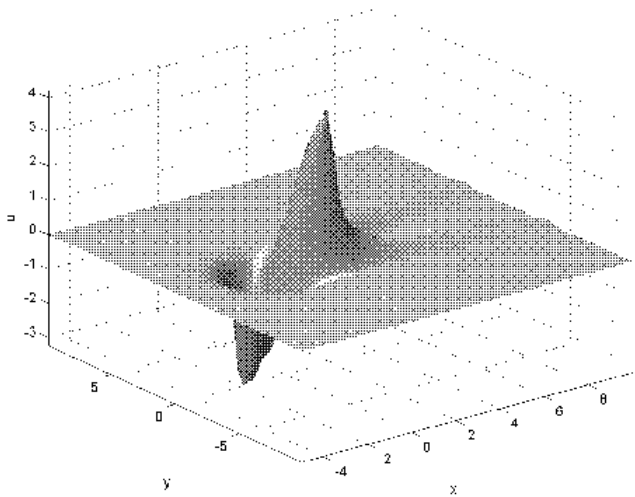}
    \caption{Solution to the regularized dKP-I equation with initial data 
    (\ref{kpld1}) and $\sigma=0.01$, plotted at time $t=0.3$.}
    \label{figdkp.3}
\end{figure}

\begin{figure}[!htb]
      \centering \includegraphics[width=10cm]{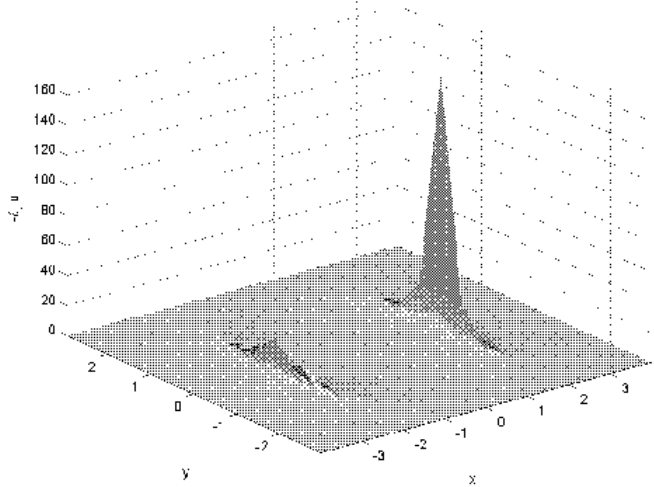}
    \caption{The $x$-derivative of the regularized dKP-I solution with initial data 
    (\ref{kpld1}) and $\sigma=0.01$, plotted at the time $t=0.3$.}
    \label{figgraddkp.3}
\end{figure}

\begin{figure}[!htb]
      \centering \includegraphics[width=10cm]{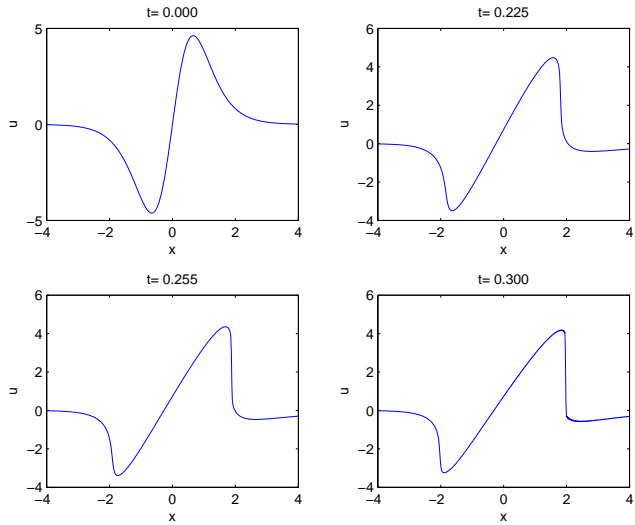}
    \caption{Solution to the regularized dKP-I equation with initial data 
    (\ref{kpld1}) and $\sigma=0.01$, plotted on the 
    $x$-axis for several values of $t$.}
    \label{figdkp4ty}
\end{figure}

To circumvent the problem of shock solutions in the formal limiting dKP model 
we shall use a \emph{dissipative regularization}. More precisely we 
add a small dissipative term, in the form 
\be
\partial_x {\left(\partial_t u + u \, \partial_x u  -\sigma\,\partial_{xx} u \right)}+ 
\lambda \, \partial_{yy} u =0, \qquad \lambda =
\pm 1.
\label{dKPreg}
\ee
with $0<\sigma \ll 1$ being some small real-valued parameter, such that $\sigma \sim \O(1)$, as $\e \to 0$. 
Equation \eqref{dKPreg} can now be seen as the KP analog of the viscous Burgers equation, \ie 
\be
\partial_t u + u \, \partial_x u  -\sigma  \,\partial_{xx} u=0.
\ee
As in (\ref{KdVint}) we obtain that (\ref{dKPreg}) can be solved via
\begin{equation}
    \partial_t\left(\E^{ t(\I\lambda k_{y}^{2}/(k_{x}+ \I \lambda
    0)+\sigma k_{x}^{2})}
    \widehat{u}\right)+\frac{\I}{2} \, k_{x}\, \E^{
    t(\I \lambda k_{y}^{2}/(k_{x}+\I \lambda0)+\sigma k_{x}^{2})}
    \, \widehat{u^{2}}= 0
    \label{dKdVint}.
\end{equation}
Obviously equation \eqref{dKPreg} is no longer conservative, but the 
dissipative term will smooth out the shocks of the dKP equation. In other words, as $\sigma\to0$, 
we expect the solution of \eqref{dKPreg} to tend to some kind of entropy solution.
\begin{remark}
    Without the dissipative regularization, the numerical code would break down shortly before the formation of shocks, 
    due to the appearing steep gradients which cause instabilities in our 
    explicit time integration scheme. However, even with an implicit, 
    unconditionally stable method, such as Crank-Nicholson, the numerical code would 
    break down close to a shock. The reason for this is mainly due to the so-called 
    \emph{aliasing error}, \ie a pollution of the spectral coefficients 
    by high frequencies, see, \eg, \cite{canuto}. Because of the   
    nonlinearity in the KP equation, this high frequency noise yields severe problems 
    near the gradient catastrophe. Even a \emph{de-aliasing}, as used in, \eg, \cite{GrKl} for the 
    KdV case, will \emph{not} stabilize the code close to the breakup, 
    since the aliasing error simply cannot be suppressed there. 
\end{remark}
Since the dKP equation is also completely integrable, many explicit 
solutions are known, \cf \cite{FeKh, Koda,KoGi}. In our study, though, we are 
interested in rather general initial data, where no closed form of the corresponding solutions is known.
In the case (\ref{kpld1}) the corresponding KP-I solution ($\lambda=-1$) is shown in Fig.~\ref{figdkp.3}, where 
$\sigma=0.01$.
The computation was carried out with $N_{x}=4096$, $N_{y}=128$, 
$L_{x}=L_{y}=10$ and $\Delta t = 5\times 10^{-5}$. Due to the dissipative 
term $\propto\sigma \partial_{xx}u$ the mass is no longer conserved, but in the given 
example the loss is only of the order of a few percent. Thus the 
shown solution should indeed be very close to a true shock solution of the dKP 
equation. Note that the tails, as $x\to \infty$, are clearly visible. 
Out focus here is now on the the region of
steep gradients in the solutions to dKP. To this end it can be seen from 
Fig.~\ref{figgraddkp.3}, that the derivative $\partial_{x}u$ is always 
\emph{maximal} on the $x$-axis (note that we plot here $-\partial_{x}u$).  
As expected the maximum of the derivative is of the order 
$1/\sigma$, and it is significantly bigger in the wave front for $x>0$, where the 
tails form. This can also be inferred from 
Fig.~\ref{figdkp4ty}, where the dKP-I solution is plotted on the $x$-axis 
for several values of the time.
We observe that the gradient catastrophe is reached for $x>0$ roughly at the time $t_c\approx  0.23$. 
For $t\geq t_c$, the gradient remains more or less unchanged due to 
the implemented dissipation in \eqref{dKPreg}. The 
gradient catastrophe for the second wave front for $x<0$ is reached roughly for $t\approx 0.3$. 

For $\lambda=1$, \ie the KP-II case, the situation is similar, but the tails are now directed towards 
$x\to-\infty$. The gradient catastrophe is now reached first in the wave front for $x<0$. 
This is due to the fact that the dKP equation \eqref{dKP}, as well as its dissipative regularization \eqref{dKPreg}, 
are \emph{invariant} under the simultaneous 
change of $x\to -x$, $u (t)\to -u(t)$, and $\lambda\to 
-\lambda$, for $t\geq 0$. For initial data of the type (\ref{kpld1}) which are odd 
in $x$, \ie $u_{I}(-x,y)=-u_{I}(x,y)$, one consequently obtains the dKP-II solution $u_+(t,x,y)$
from the dKP-I solution $u_-(t,x,y)$, with the same initial data, by identifying $u_{+}(t,x,y)=-u_{-}(t,-x,y)$. 

\subsection{KP oscillations} 
In the following we shall study the oscillations for small, but still non-zero, 
$\e\ll 1$ in the solution of \eqref{KP}. 

In the closely related KdV case this limiting regime is rather well understood, see, \eg, \cite{GT, LL, Le, FRT1}: 
Roughly speaking, the corresponding analytical approach consists 
of the following steps: After the breaking in the corresponding Hopf 
solution \eqref{hopf}, an oscillatory zone is identified via the solution of the 
so-called \emph{Whitham equations}, which depend only on the slow coordinates 
$x$ and $t$. The corresponding approximate solution to the KdV 
equation in this region is consequently given in terms of theta 
functions, evaluated 
on the fast coordinates $X=x/\epsilon$, $T=t/\epsilon$, whereas the branch points 
of the underlying Riemann surface are determined via Whitham's modulation equations. 
Outside the oscillatory zone, the KdV solution is 
approximated by the corresponding solution of the Hopf equation. For 
a comparison of this asymptotic solution with a numerical KdV 
solution see \cite{GrKl}. 

There are several obstacles to overcome when one tries an analogous approach in the KP case. 
First, one should keep in mind that the formal dispersionless KdV model, 
\ie the Burger's equation \eqref{Bu}, is a first order PDE which can be integrated by the method of 
characteristics. On the other hand, the corresponding dKP equation, 
is a second order PDE. Although the dKP equation is completely 
integrable and thus explicit solutions are known, there exists no general procedure 
of integrating this equation for generic initial data so far. 
Secondly, the Whitham equations in the KdV case can be brought, via the hodograph transform, into the so-called 
Riemann linear form. They can then be again solved (at least implicitly) for generic 
initial data. Thus all quantities entering the theta functional solution, 
as for instance the phase, are known explicitly via their dependence on the 
branch points of the Riemann surface defined by the solution of the 
Whitham equations. 

\begin{remark} In the KP case, the corresponding Whitham equations 
were solved formally in \cite{Kr}. So far, however, it is not clear how to 
bring this solution into a form which is convenient for a numerical 
simulation. This is especially true for the non-monotonous 
initial data with two inflection points we are considering here, a 
case which has not even been studied in the much better understood 
KdV setting. (Note that the two inflection points are needed to satisfy the 
constraint (\ref{const}).) In addition it would be necessary to have an explicit 
formula for the corresponding phase entering the theta function to be 
able to compare the asymptotic solution to the exact solution in the 
Whitham zone.
\end{remark}
We hope that the numerical results presented in this paper help to 
develop a similar asymptotic description as in the KdV case which 
will be the subject of further research. 

Now, we shall first show (numerically) that the solution of the dKP equation gives the 
correct limiting behavior as $\epsilon\to 0$ before breakup. To this end 
we compare the dKP-I solution $u_{\rm dKP}(t)$, with initial data (\ref{kpld1}), 
and the corresponding KP-I solution $u_{\rm KP}(t)$ at times $t\leq 0.2$, \ie 
up to times close to (but still before) the appearance of the first shock in the 
dKP-I solution. We shall consider values of $\epsilon$ between $0.1$ and $0.01$, with 
$N_{x}=2^{11}$, $N_{y}=2^{7}$, and $\Delta t=2\times 10^{-5}$. 
Note that we solve here the dKP equation without dissipative 
regularization since we stop the computation before the appearance of the 
gradient catastrophe. We also remark that this is possible with 
a relative mass conservation of $10^{-4.39}$. 

\begin{figure}[!htb]
      \centering \includegraphics[width=10cm]{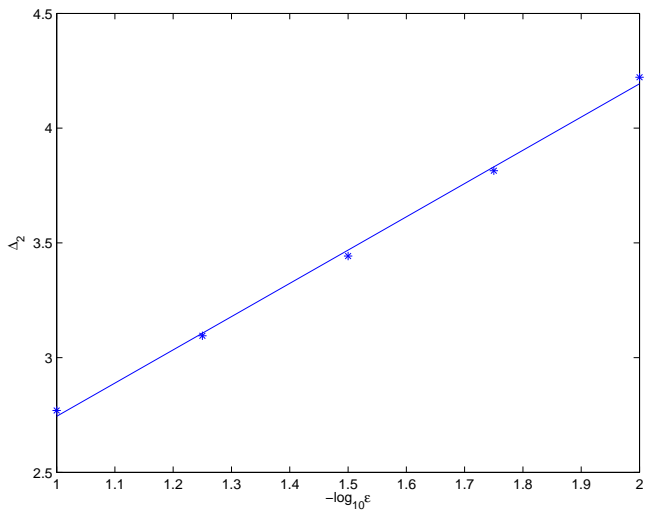}
    \caption{Error $\Delta_{2}$ for several 
    values of $\epsilon$. The data can be fitted by least square 
    analysis with a straight line 
    $-\log_{10}\Delta_{2}=-a\log_{10}\epsilon+b $ with $a=1.45$ and 
    $b=1.30$. 
    The correlation coefficient is $r=0.999$, the standard error for $a$ 
    is $\sigma_{a}=0.056$.}
	\label{figdelta2}
\end{figure}

\begin{figure}[!htb]
      \centering \includegraphics[width=10cm]{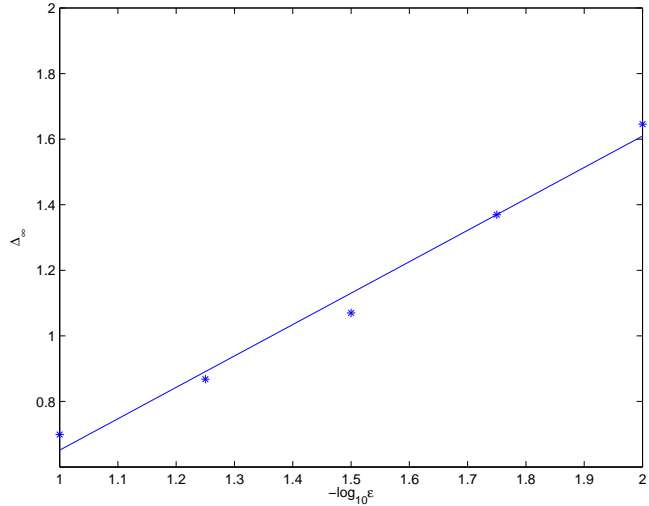}
    \caption{Error $\Delta_{\infty}$ for several 
    values of $\epsilon$. The data can be fitted by least square 
    analysis with a straight line 
    $-\log_{10}\Delta_{\infty}=-a\log_{10}\epsilon+b $ with $a=0.96$ and 
    $b=-0.31$. 
    The correlation coefficient is $r=0.993$, the standard error for $a$ 
    is $\sigma_{a}=0.099$.}
	\label{figdeltai}
\end{figure}

We find that the difference between the solutions $u_{\rm dKP}(t)$, $t<t_c$, and $u_{\rm KP}(t)$ 
increases monotonically with time. In the considered range of 
$\epsilon$ and for $t=0.2$, the $L^{2}$ norm difference $\Delta_{2}$, defined in \eqref{delta}, 
decreases roughly like $\O(\epsilon^{3/2})$, as $\e\to 0$. 
More precisely we find that the data can be fitted by a straight line 
$-\log_{10}\Delta_{2}=-a\log_{10}\epsilon+b $, with $a=1.45$ and 
$b=1.30$, as can be seen in Fig.~\ref{figdelta2}. 
The correlation coefficient is found to be $r=0.999$, the standard error for $a$ 
is $\sigma_{a}=0.056$. 
\begin{remark} Note that the $\O(\e^{3/2})$ error in $\Delta_{2}$ fits with our 
results of Section \ref{smod}. There we numerically found that the $\Delta_{2}$ error for 
asymptotically small solutions, \ie solutions of the order $\O(\e)$, decays like $\O(\e^{5/2})$ for 
small $\e\ll1$. 
\end{remark} 
For the corresponding $L^{\infty}$ norm difference $\Delta_{\infty}$ we find numerically that the error roughly decreases like  
$\epsilon$. More precisely again, we get (see Fig.~\ref{figdeltai}), that the 
data can be fitted by a straight line 
$-\log_{10}\Delta_{2}=-a\log_{10}\epsilon+b $ with $a=.96$ and 
$b=-0.31$. 
The correlation coefficient is found to be $r=0.993$, the standard error for $a$ 
is $\sigma_{a}=0.099$. 
Notice that at $t=0.2$, the first oscillations of the KP solutions 
appear as can be seen below. This leads to considerably large values of 
$\Delta_{\infty}$ since the solution to the dKP equation will never 
show oscillations.

For times $t\geq t_c$, \ie beyond the onset-time of the first gradient catastrophe, 
we shall study the KP-I solution, corresponding to the initial condition (\ref{kpld1}), 
for $0.01\leq \epsilon \leq 0.1$ and times $t\leq 0.4$. The 
computations are carried out with $N_{x}=2048$ to $N_{x}=8192$, 
$N_{y}=128$, $L_{x}=L_{y}=5$, and $\Delta t=2\times10^{-5}$ to $\Delta 
t=4\times10^{-6}$. Thereby we observe a relative mass conservation of the order of 
$10^{-4}$ to $10^{-3}$. 
The following observations are now in order:
\begin{itemize}
\item 
It can be seen in Fig.~\ref{figkp1e2.4} for the KP-I case that oscillations mainly form in the 
region of large (spatial) gradients appearing in the dKP solution given above. 
\begin{figure}[!htb]
      \centering \includegraphics[width=10cm]{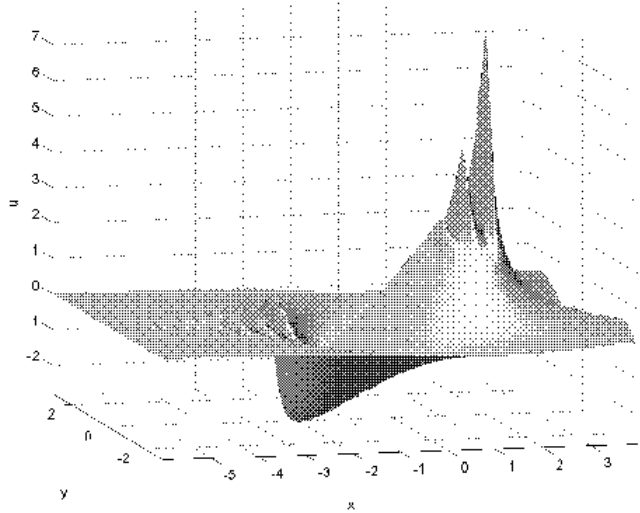}
    \caption{Solution of the KP-I equation obtained from the initial 
    data (\ref{kpld1}) for $\epsilon=0.1$, at time 
    $t=0.4$.}
    \label{figkp1e2.4}
\end{figure}
\item As in the KdV case the number of oscillations increases with decreasing 
$\epsilon$, whereas the wavelength decreases, see Fig.~\ref{figkp1e4.4}.
\begin{figure}[!htb]
      \centering \includegraphics[width=10cm]{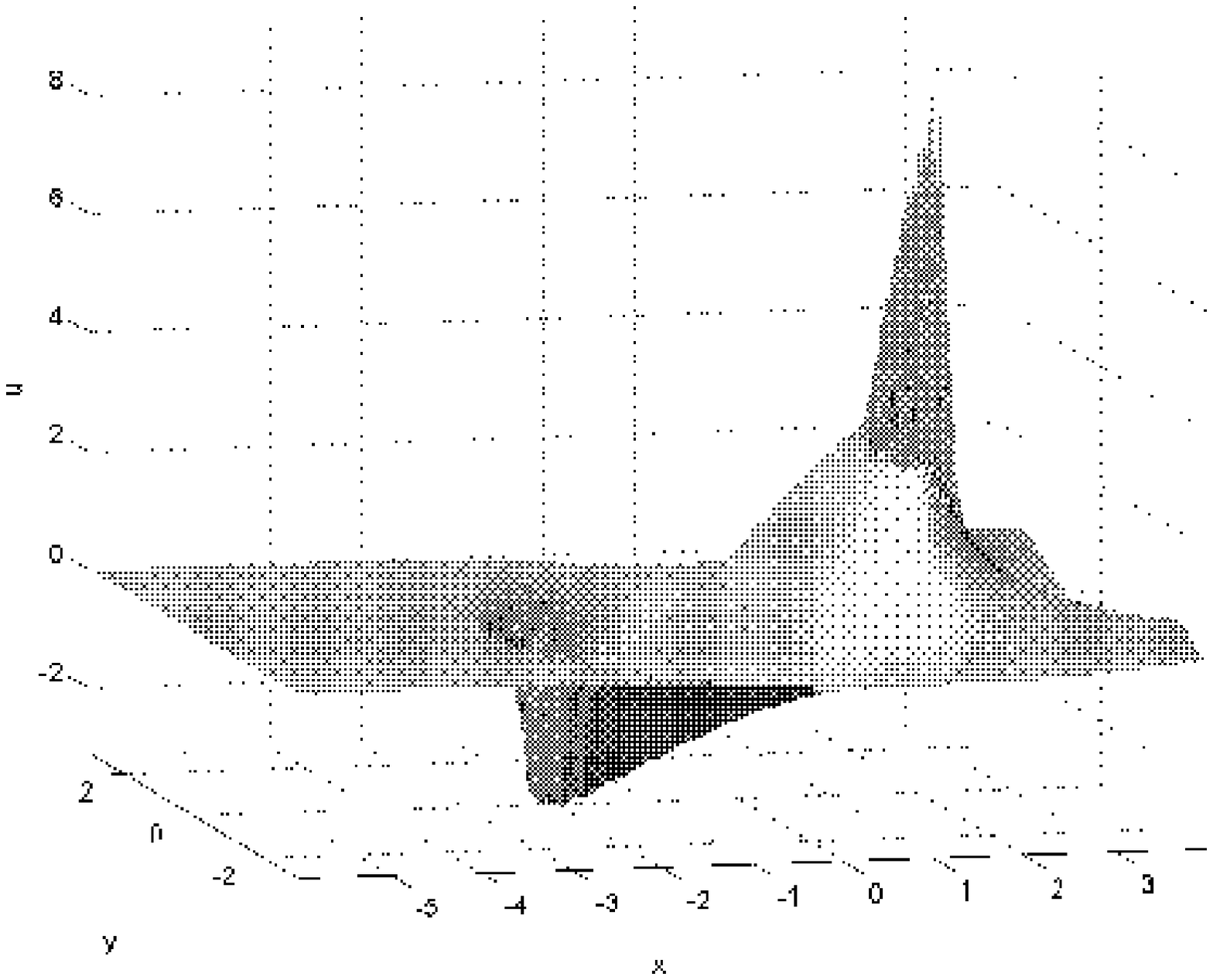}
    \caption{Solution of the KP-I equation obtained from the initial 
    data (\ref{kpld1}) for $\epsilon=0.01$, at time 
    $t=0.4$.}
    \label{figkp1e4.4}
\end{figure}
\item As expected there are \emph{two oscillatory regions}, one for $x<0$ 
and one for $x>0$, both of which are shown in detail in 
Fig.~\ref{figkp1e4.42}.
\begin{figure}[!htb]
      \centering \includegraphics[width=14cm]{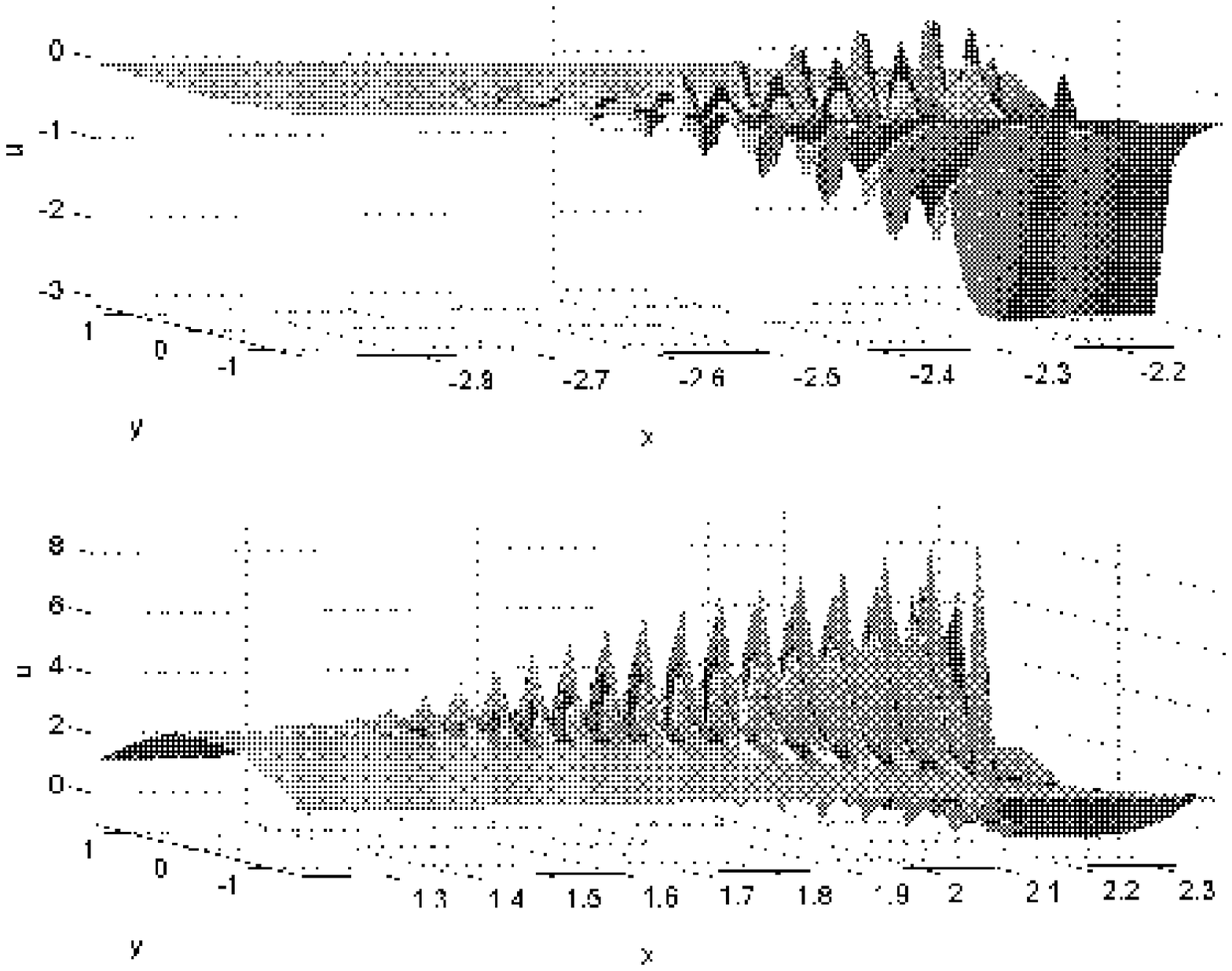}
    \caption{Solution of the KP-I equation obtained from the initial 
    data (\ref{kpld1}) for $\epsilon=0.01$, at time 
    $t=0.4$.}
    \label{figkp1e4.42}
\end{figure}
To get a complementary view of the rapid oscillations in this region 
a contour plot of the same situation is shown in Fig.~\ref{figkp1e4.42c}.
\begin{figure}[!htb]
      \centering \includegraphics[width=14cm]{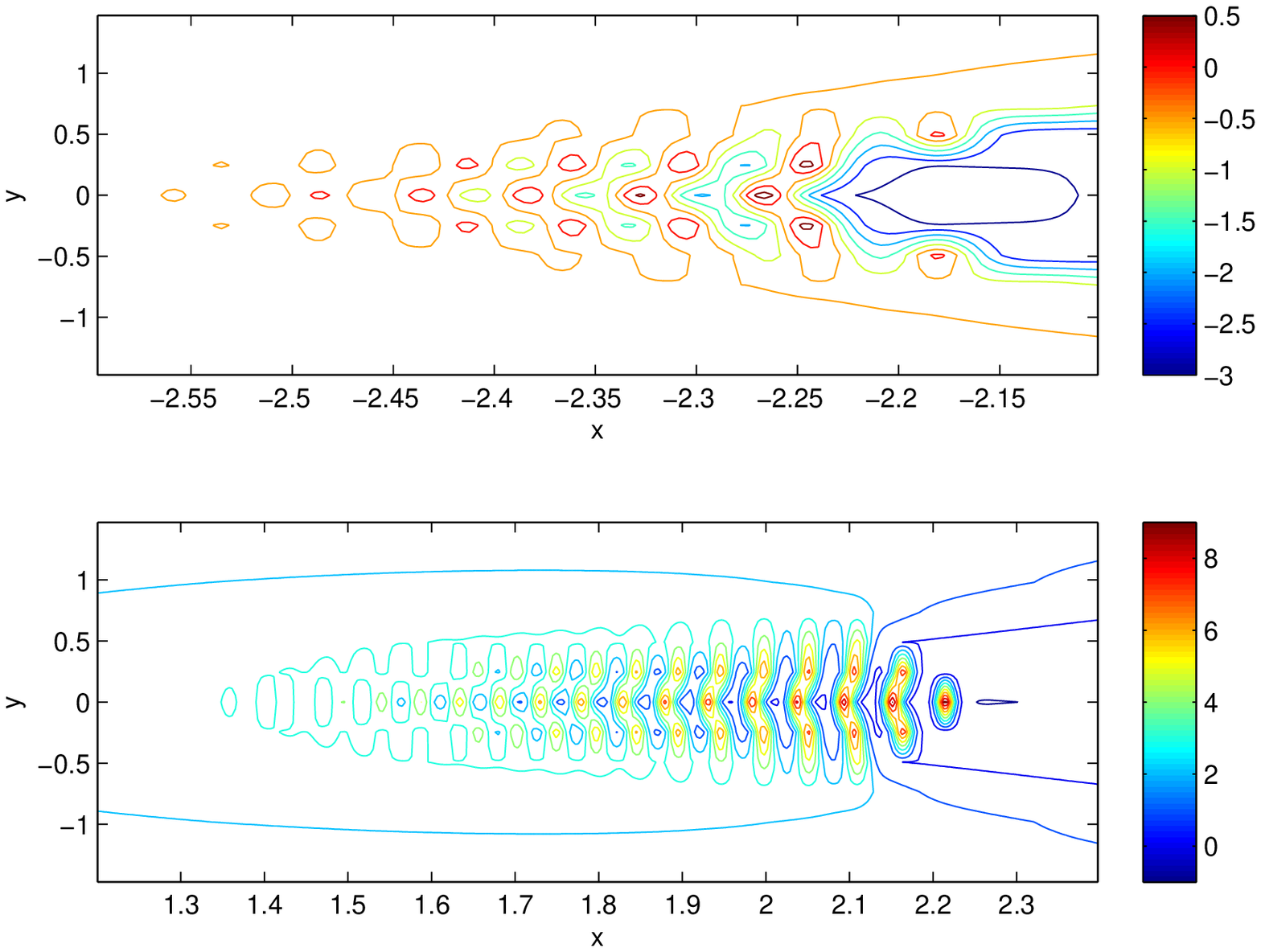}
    \caption{Solution of the KP-I equation obtained from the initial 
    data (\ref{kpld1}) for $\epsilon=0.01$, at time 
    $t=0.4$}
    \label{figkp1e4.42c}
\end{figure}
\item The oscillations are always at \emph{most rapid on the $x$-axis}, see 
Fig.~\ref{figkp1e4y.4} and the results of the next subsection.
\begin{figure}[!htb]
      \centering \includegraphics[width=10cm]{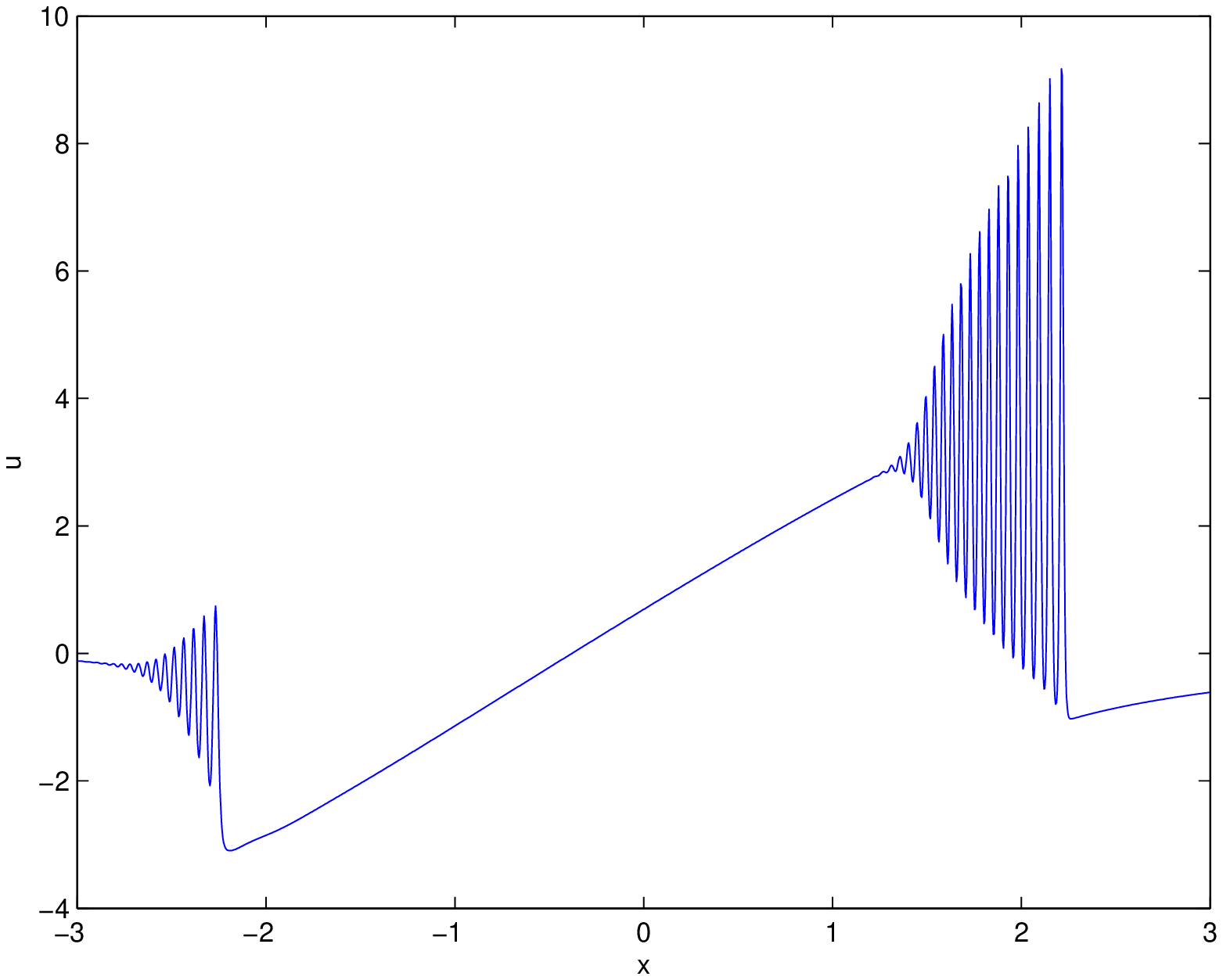}
    \caption{Solution of the KP-I equation obtained from the initial 
    data (\ref{kpld1}) for $\epsilon=0.01$, at time 
    $t=0.4$ and $y=0$.}
    \label{figkp1e4y.4}
\end{figure}
\end{itemize}
We note that the fast oscillations are numerically well resolved, 
as can be inferred from Fig.~\ref{figkp1e4y.42}, which shows two enlarged sections of the oscillatory zone.
\begin{figure}[!htb]
      \centering \includegraphics[width=10cm]{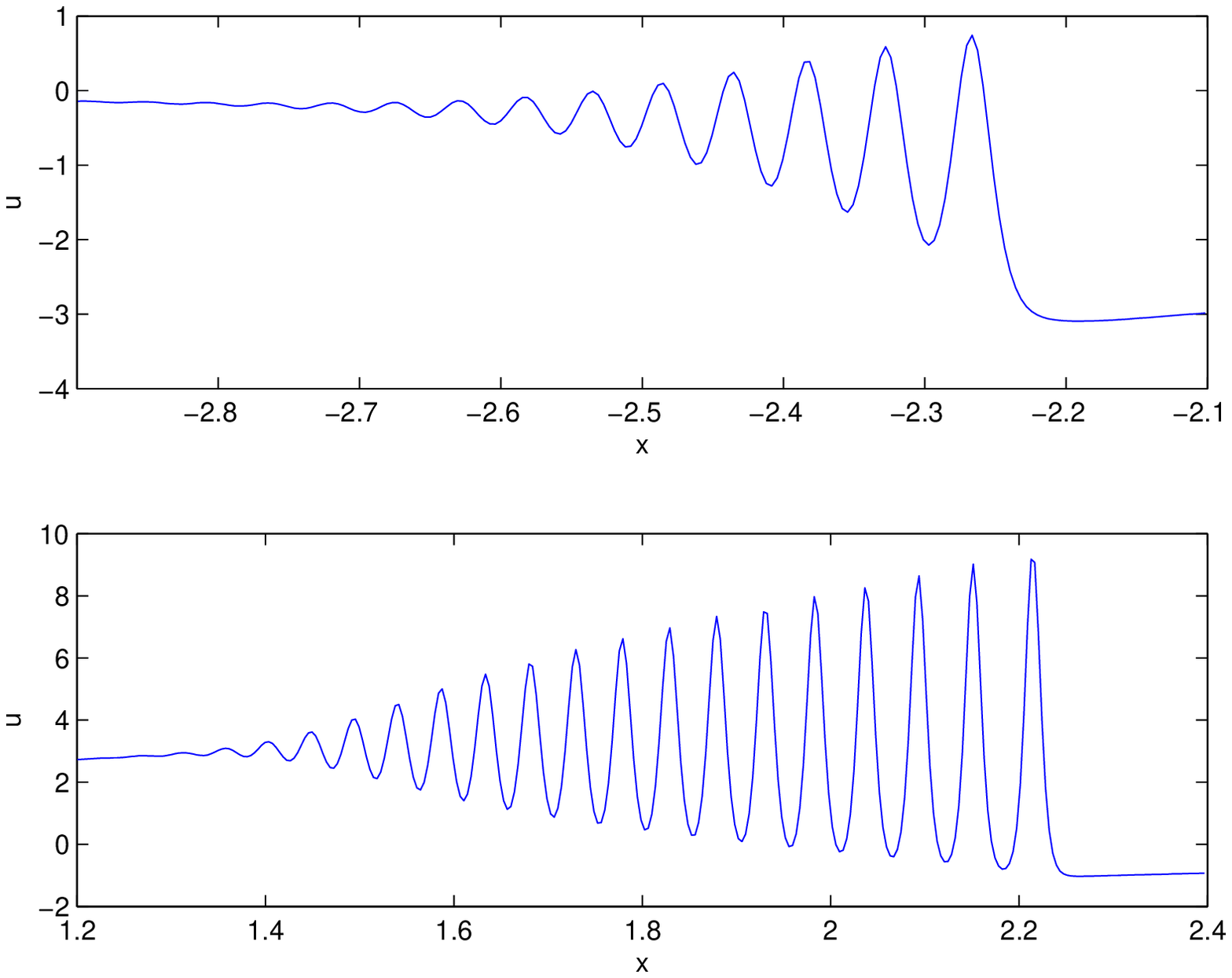}
    \caption{Solution to the KP-I equation obtained from the initial 
    data (\ref{kpld1}) for $\epsilon=0.01$, at time 
    $t=0.4$ and $y=0$.}
    \label{figkp1e4y.42}
\end{figure}
Moreover we observe that the time evolution of the initial data \eqref{kpld1} is as expected from the 
analysis of the regularized dKP equation above. 

\begin{figure}[!htb]
      \centering \includegraphics[width=14cm]{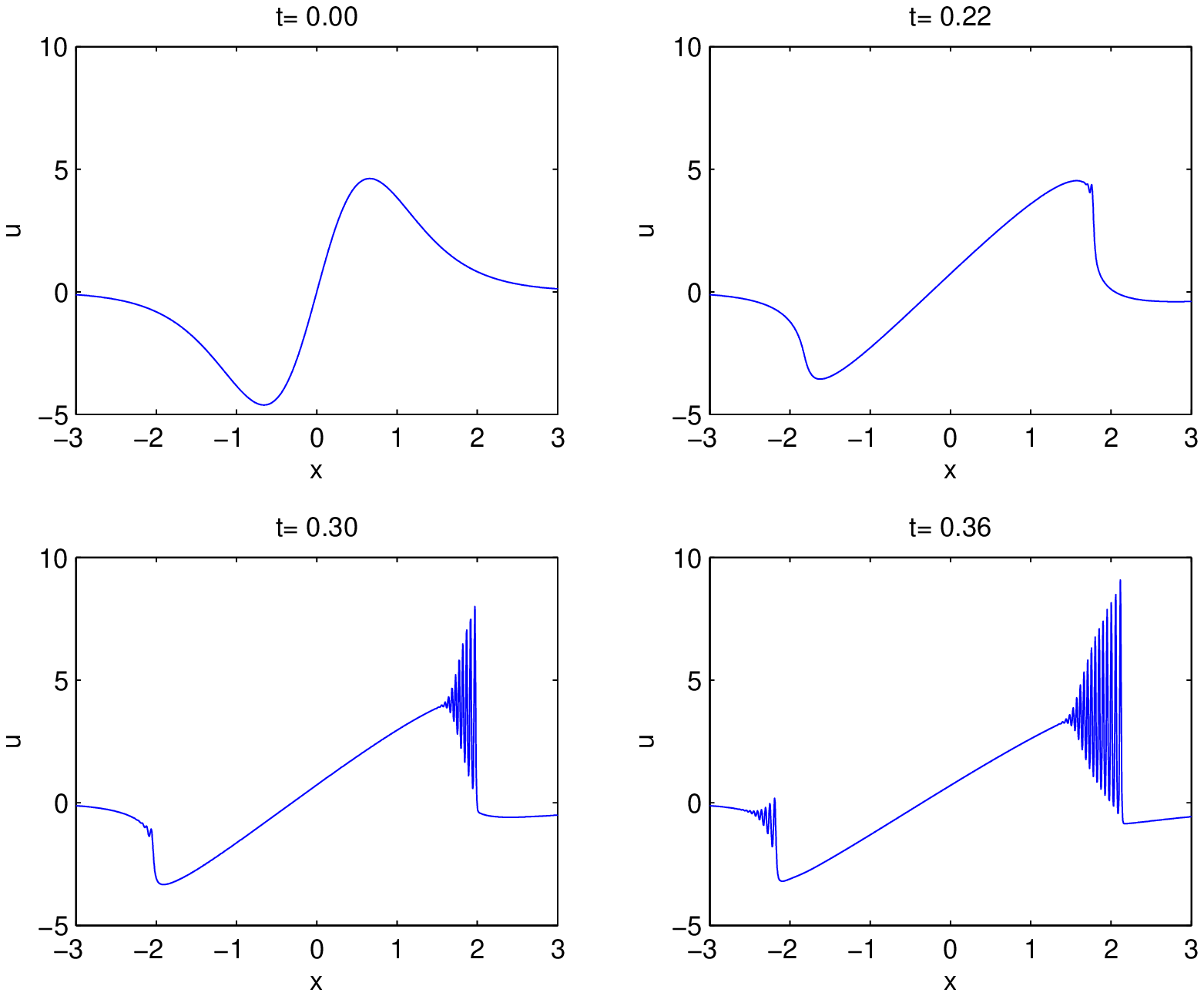}
    \caption{Solution of the KP-I equation obtained from the initial 
    data (\ref{kpld1}) for $\epsilon=0.01$ for several 
    values of the time.}
    \label{figkp1e4y4t}
\end{figure}

\begin{figure}[!htb]
      \centering \includegraphics[width=14cm]{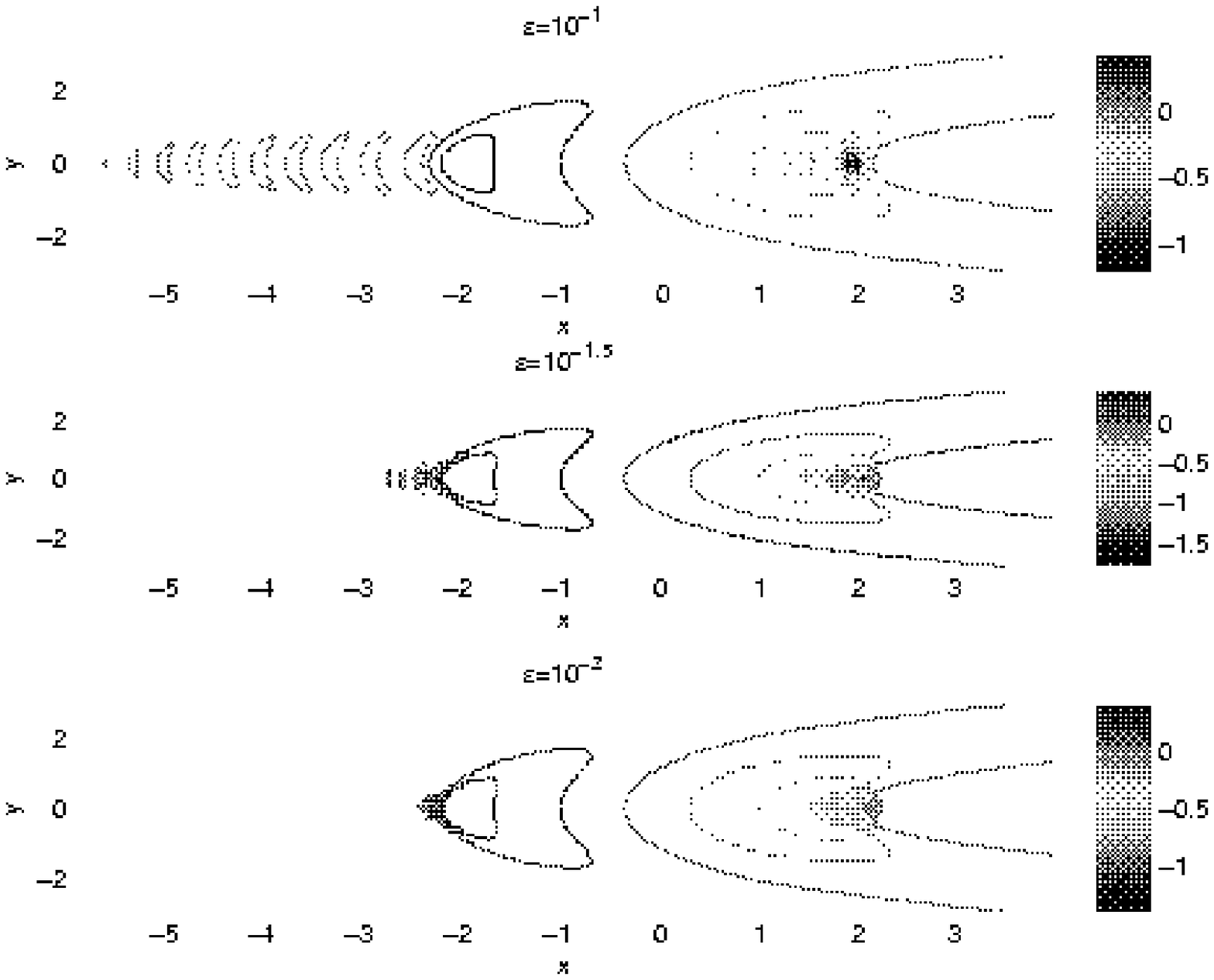}
    \caption{Solution of the KP-I equation obtained from the initial 
    data (\ref{kpld1}) at time 
    $t=0.4$ for several values of $\epsilon$.}
    \label{figkp3e}
\end{figure}

\begin{figure}[!htb]
      \centering \includegraphics[width=14cm]{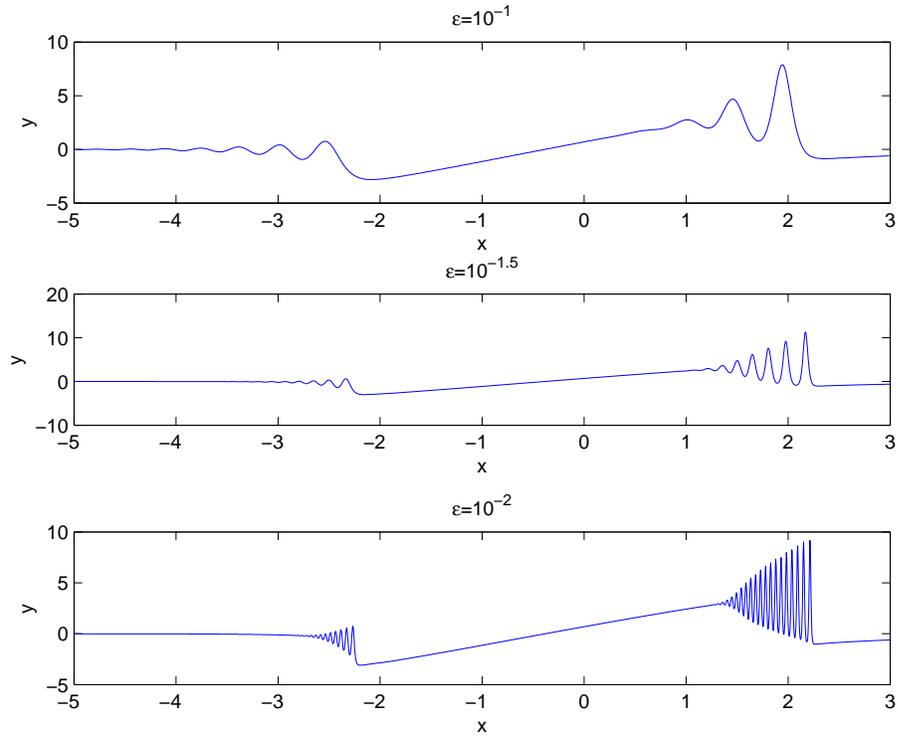}
    \caption{Solution of the KP-I equation obtained from the initial 
    data (\ref{kpld1}) for several values of $\epsilon$, at time 
    $t=0.4$ and $y=0$.}
    \label{figkp3ey}
\end{figure}

\begin{figure}[!htb]
      \centering \includegraphics[width=10cm]{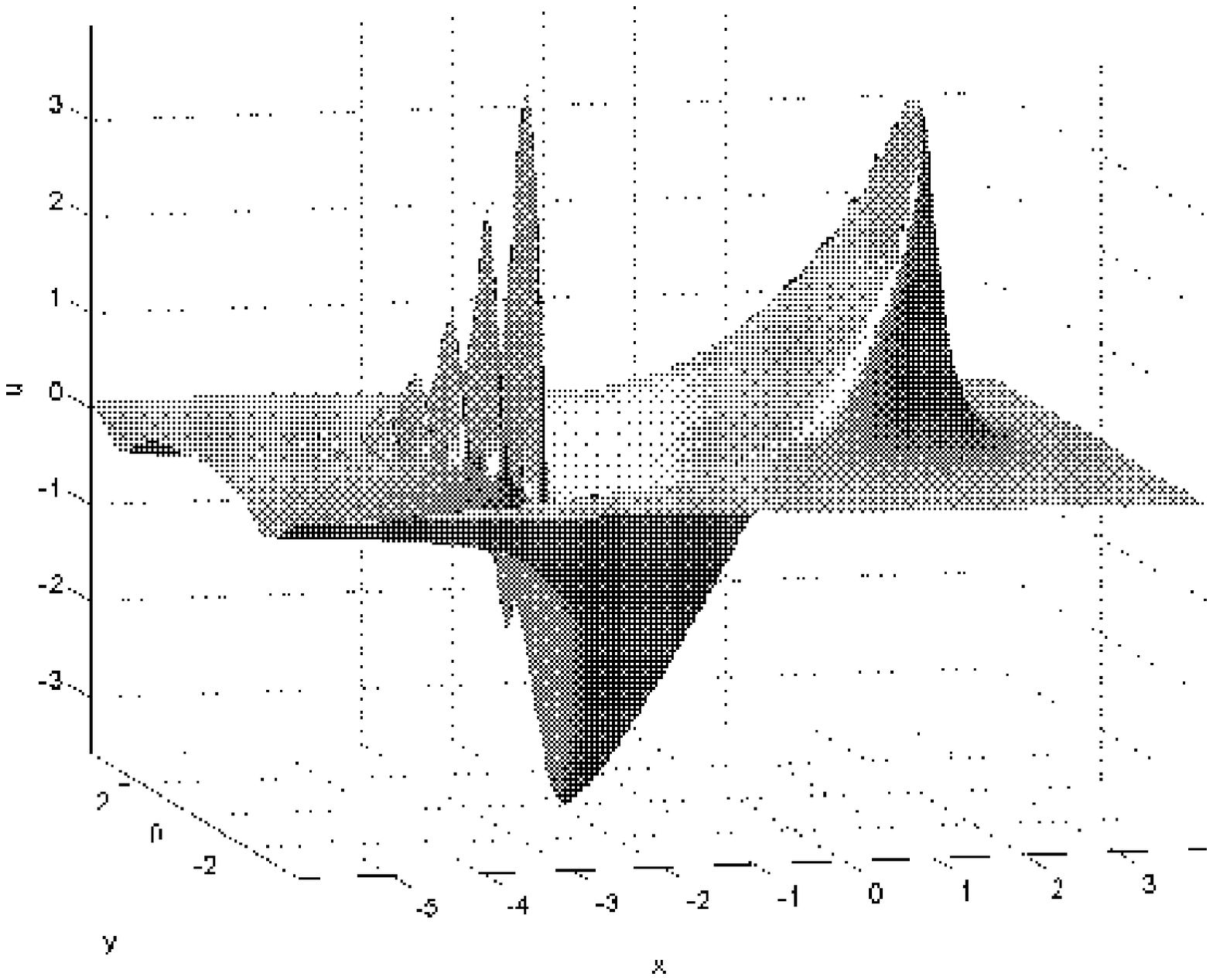}
    \caption{Solution of the KP-II equation obtained from the initial 
    data (\ref{kpld1}) for $\epsilon=0.1$ at time 
    $t=0.4$.}
    \label{figkp1e2m}
\end{figure}

\begin{figure}[!htb]
      \centering \includegraphics[width=10cm]{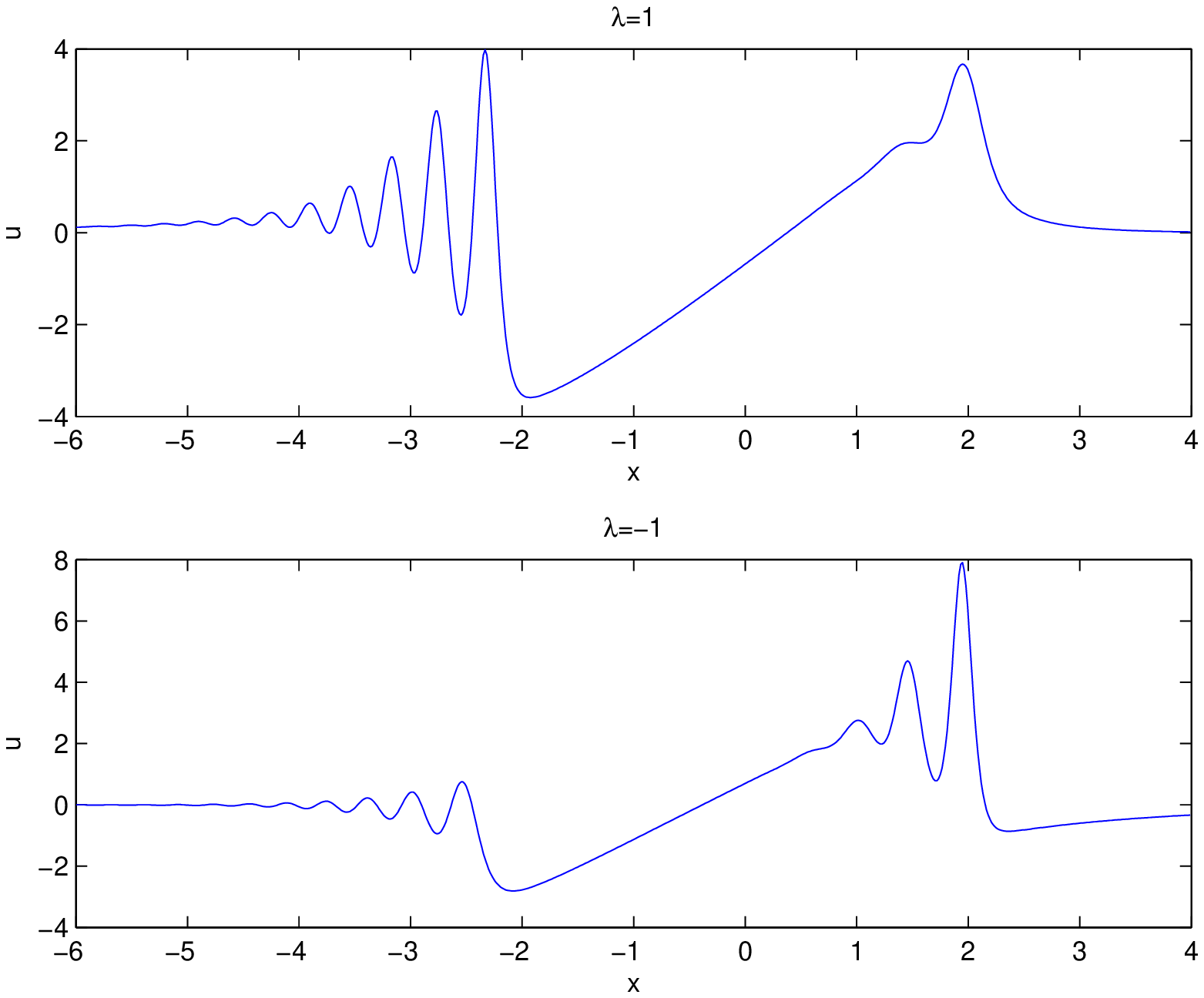}
    \caption{Solutions to both KP models obtained from the same initial 
    data (\ref{kpld1}), shown at time 
    $t=0.4$ and $y=0$.}
    \label{figkp1e2y2l}
\end{figure}

To obtain more insight, we shall focus in Fig.~\ref{figkp1e4y4t} on the KP-I solution plotted at $y=0$, \ie 
where the rapid oscillations appear first and where they are the most pronounced. 
We note that the first oscillation appears at the time $t\approx 0.22$, for positive values of $x\in \R$, 
\ie shortly before reaching the gradient catastrophe of the dKP equation. On the other hand at time $t\approx 0.3$, 
the first oscillation appears at a negative value of the $x$-axis. Again this is shortly before a shock 
in the second wave front is reached. Both regions then develop more and 
more oscillations as time goes on until finally the situation described above and shown 
in Fig.~\ref{figkp1e4y.4} and Fig.~\ref{figkp1e4y.42} is reached.
In Fig.~\ref{figkp1e4.42} and Fig.~\ref{figkp1e4.42c} 
one can also see the $y$-dependence of the observed oscillations. 
One recognizes that although they are mainly concentrated on the 
$x$-axis, there are some oscillations for small $0<|y|\ll 1$ 
(see also the results of the next subsection). 
\begin{remark}
In the corresponding situation for the KdV equation, the oscillatory zone \emph{shrinks} with 
$\epsilon$. The same is the case for the KP model as can be 
inferred from  Fig.~\ref{figkp3e} and Fig.~\ref{figkp3ey}. Note that this shrinking is also present in the $y$-direction.
\end{remark} 
In contrast to the dKP equation, the KP equation is \emph{not} invariant 
under the transformation $x\to -x$, $u (t)\to -u(t)$, and $\lambda\to 
-\lambda$. Thus the rapid oscillations for small $\e \ll 1$ in the KP-II 
equation will be rather different from the KP-I case, as can be seen from Fig.~\ref{figkp1e2m}.
Here, the tails directed to $x\to-\infty$ are clearly visible as well as the strong 
oscillations for $x<0$. On the other hand there are \emph{virtually no oscillations} for $x>0$. 
For the sake of comparison we plot in Fig.~\ref{figkp1e2y2l}, the situation 
for both KP models, \ie for the cases $\lambda=\pm1$, at $y=0$ and $t=0.4$. 
Whereas there are rather strong oscillations near \emph{both} breaking wave fronts in the 
KP-I case ($\lambda=-1$), the KP-II case ($\lambda=1$) enhances oscillations for $x<0$ 
while for for $x>0$ they are more or less suppressed. 

\subsection{Dependence on the $y$-coordinate}
As already mentioned, the KP equation was originally derived to 
describe quasi one-dimensional (dispersive) wave phenomena with only weak 
transverse effects. Thus 
physically interesting solutions will  have a rather weak dependence on 
the coordinate $y\in \R$. Mathematically it is nevertheless interesting 
to study the effects of a stronger $y$-dependence. In this subsection 
we will thus have a closer look on this dependence in the context 
of low-dispersion. 

\begin{figure}[!htb]
    \centering \includegraphics[width=10cm]{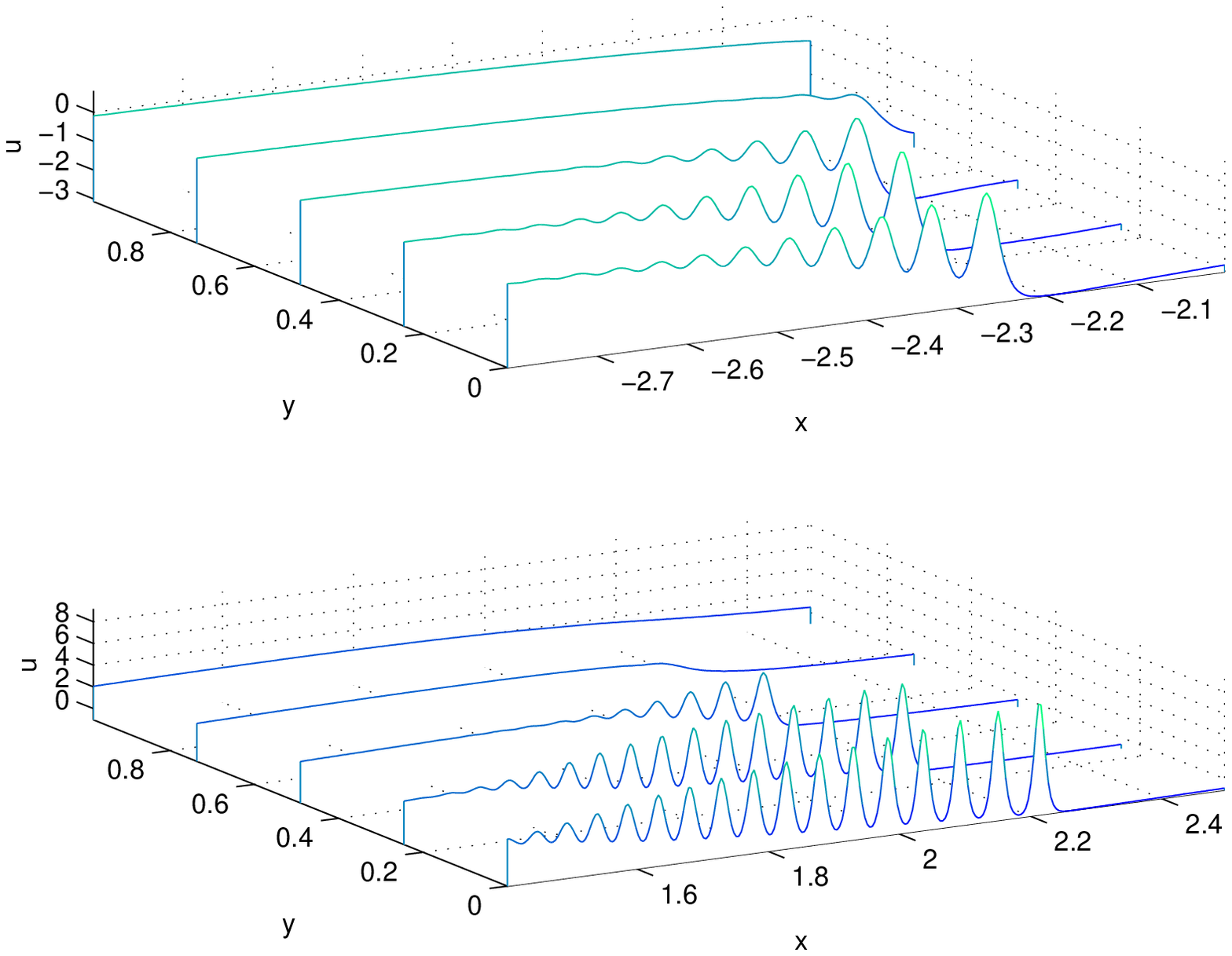}
    \caption{Solution of the KP-I equation obtained from the initial 
    data (\ref{kpld1}) for $\epsilon=0.01$ at time 
    $t=0.4$.}
    \label{figkp1e42w}
\end{figure}

Recalling the analysis of the Green's function for the linear part of the KP 
equation, given in Section 2.1, it appears that only the 
Airy-part leads to oscillations, whereas the term $\propto \partial_{yy} u$ 
is responsible for the formation of tails. 
On the other hand, the gradient catastrophe in the dKP model is 
the result of the nonlinear term $\propto u\partial_{x}u$, as in the case of the Hopf equation. Consequently one 
expects for initial data with a non-trivial $y$-dependence that oscillations 
can only appear in a small vicinity of the $x$-axis, \ie before 
the formation of tails takes over. This behavior has already been observed in the subsection above and 
it is further supported by Fig.~\ref{figkp1e42w} which shows the KP solution for $\epsilon=0.01$ 
and $t=0.4$ in the vicinity of the $x$-axis.
To study the effect of the dependence on the $y$-coordinate in 
more detail, we consider again initial data of the form (\ref{kpld1}). However we shall 
now evaluate them  at $R=R_\nu$, given by 
\begin{equation}
    R_\nu^{2}=x^{2}+\nu y^{2}
    \label{Rn},
\end{equation}
where now $\nu\in \mathbb{R}_{0}^{+}$ is a \emph{deformation parameter}. Obviously, for 
$\nu =0$, there is no $y$-dependence, and the KP model reduces to 
the KdV equation. The parameter $\nu$ thus allows for a continuous 
deformation of the KdV sector. 

\begin{figure}[!htb]
    \centering \includegraphics[width=10cm]{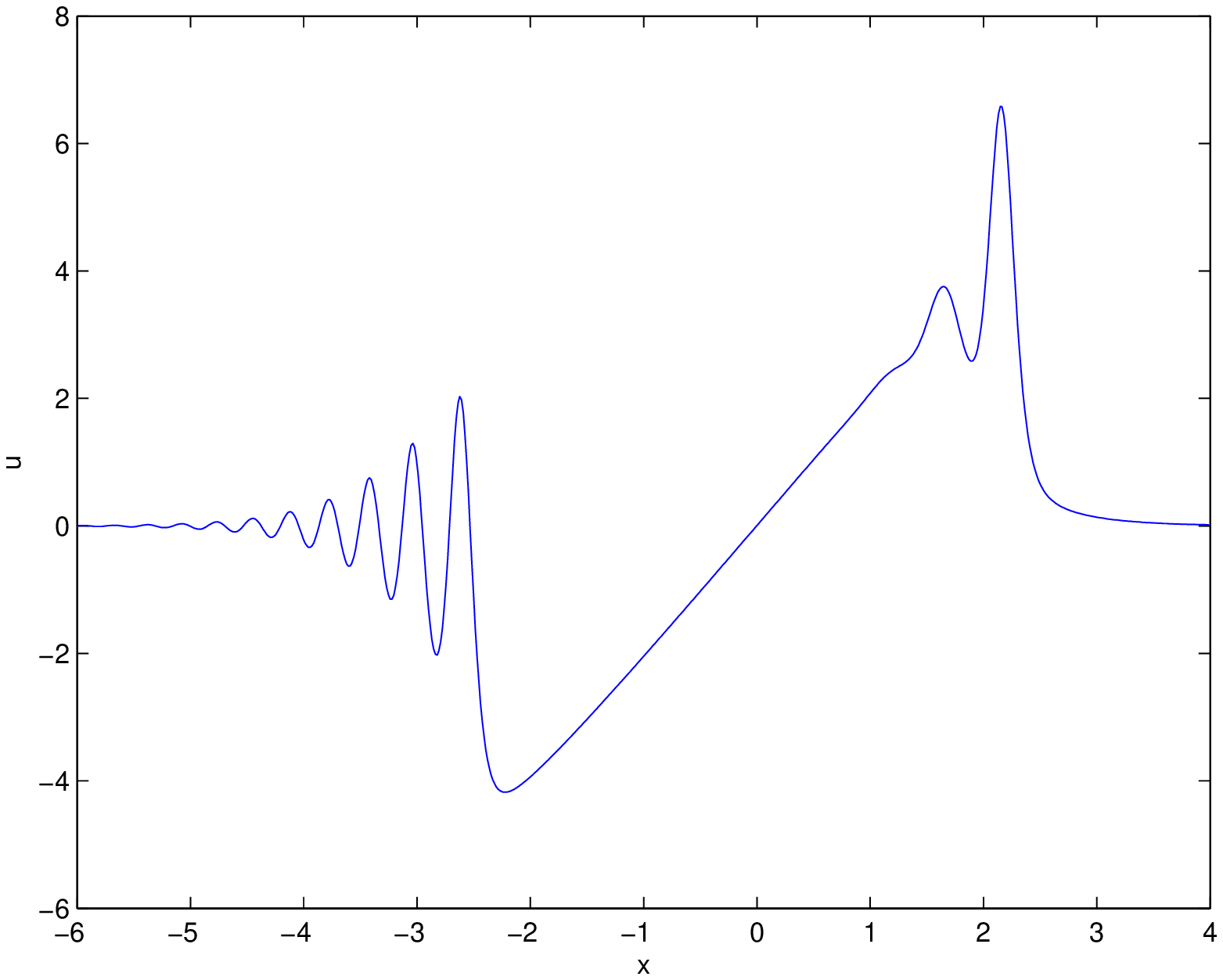}
    \caption{Solution of the KdV equation with $\epsilon=0.1$ obtained from the initial 
    data (\ref{kpld1}) with (\ref{Rn}) and $\nu=0$ for the time 
    $t=0.4$.}
    \label{figkdv1e2der}
\end{figure}

\begin{figure}[!htb]
    \centering \includegraphics[width=14cm]{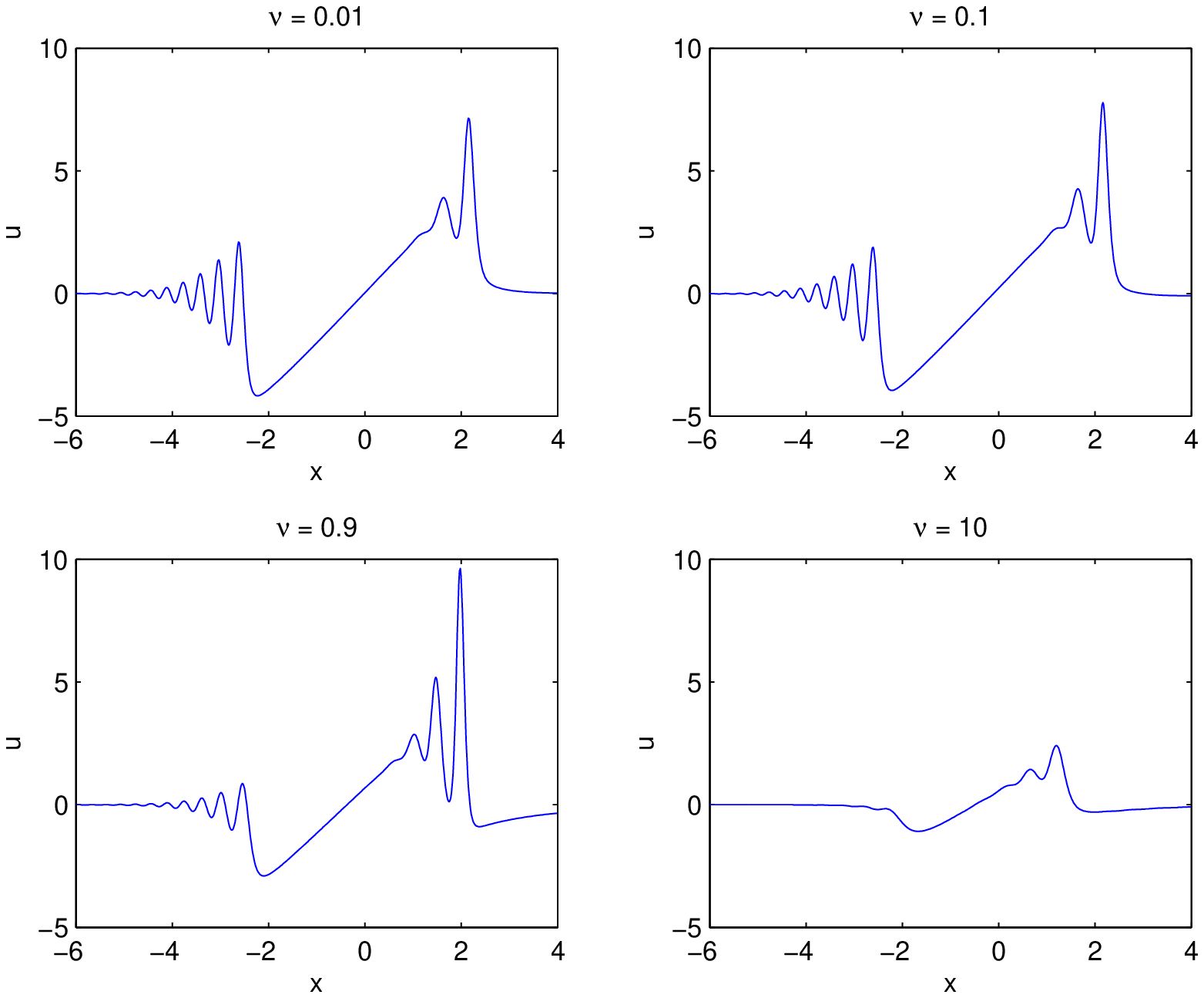}
    \caption{Solution of the KP-I equation obtained from the initial 
    data (\ref{kpld1}), (\ref{Rn}) for several 
    values of $\nu$ at time $t=0.4$.}
    \label{figkp1e24nu}
\end{figure}

\begin{figure}[!htb]
    \centering \includegraphics[width=14cm]{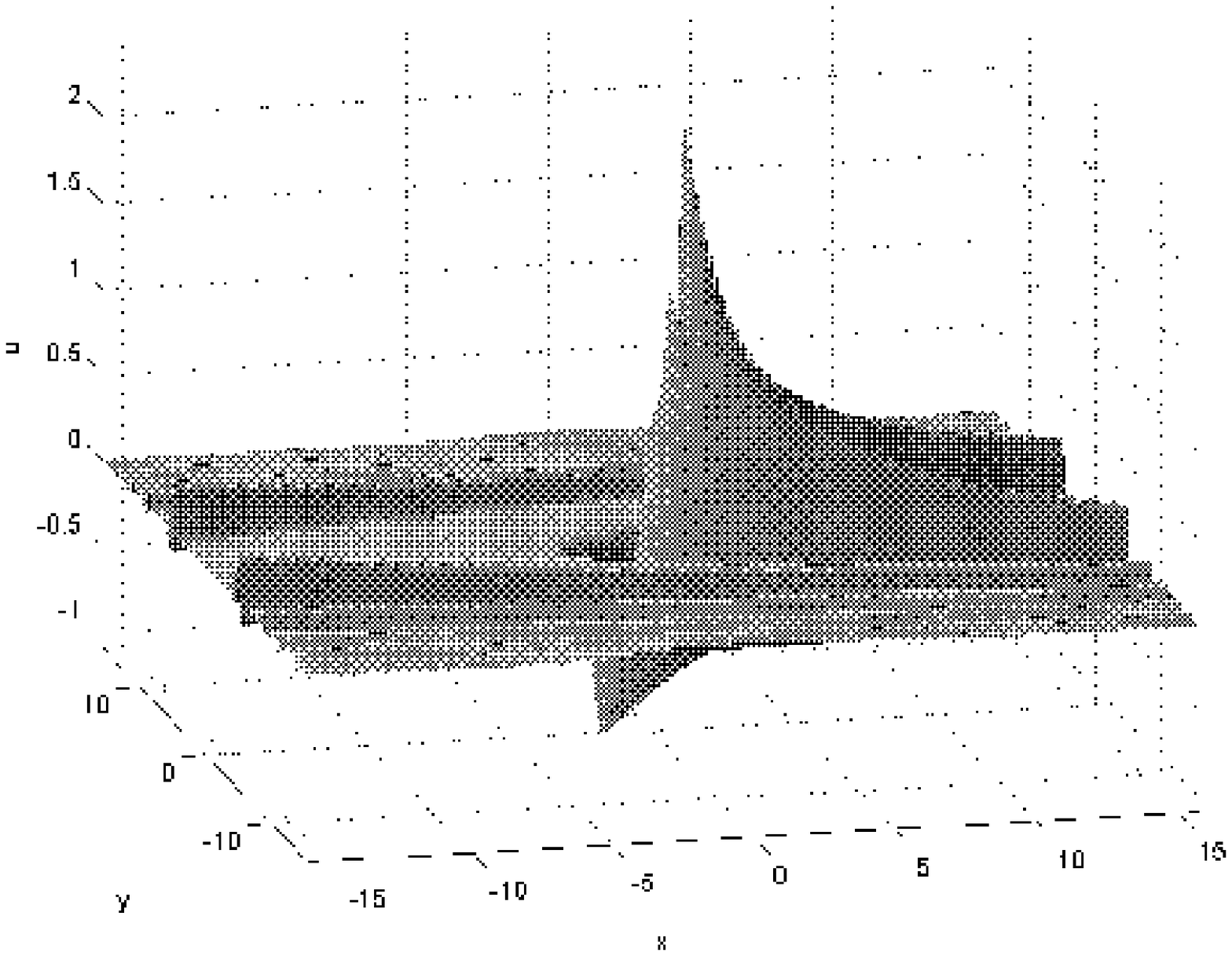}
    \caption{Solution of the KP-I equation obtained from the initial 
    data (\ref{kpld1}) with (\ref{Rn}) and $\nu=10$ at time 
    $t=0.4$.}
    \label{figkp1e2nu10}
\end{figure}

\begin{figure}[!htb]
    \centering \includegraphics[width=14cm]{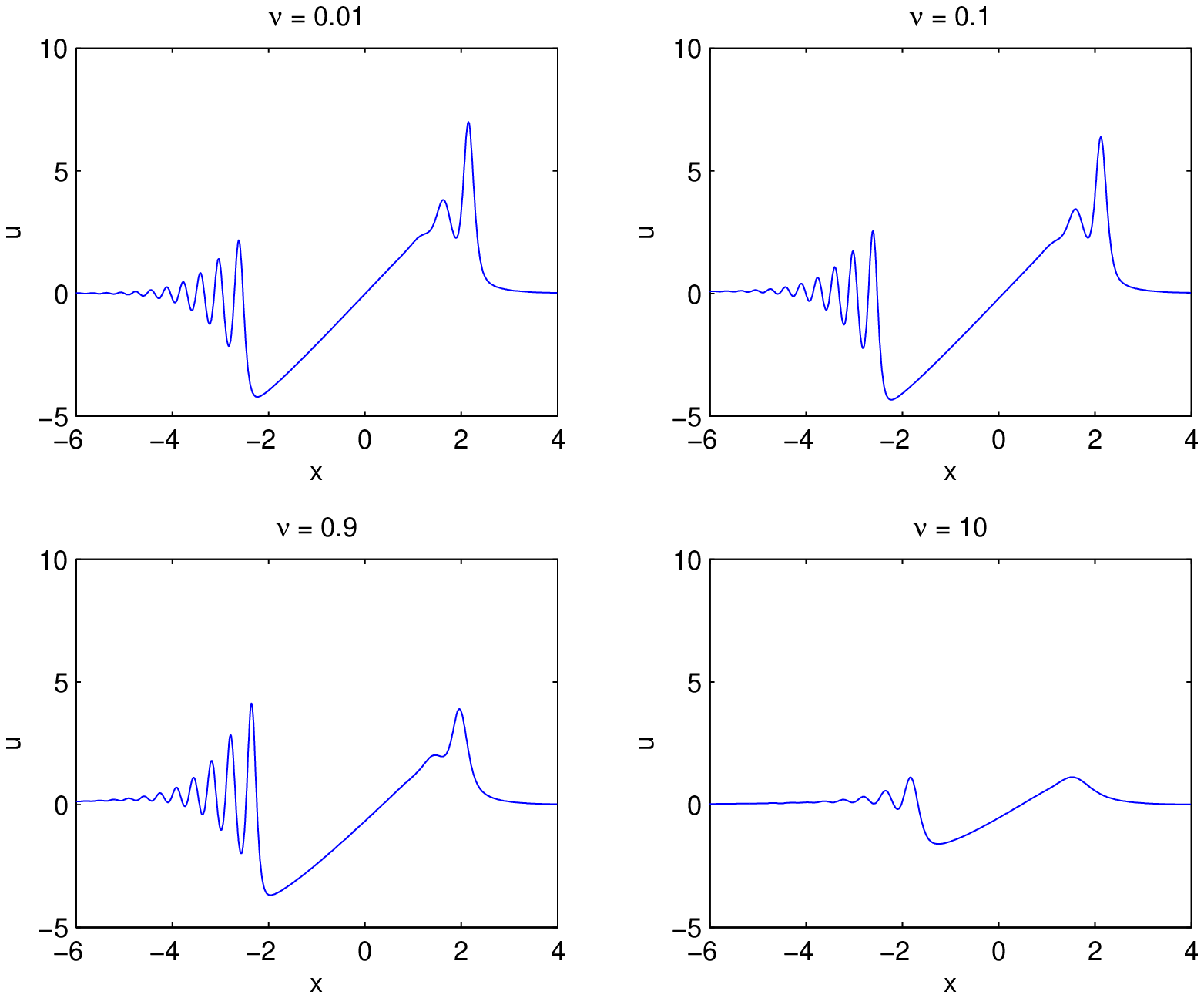}
    \caption{Solution of the KP-II equation obtained from the initial 
    data (\ref{kpld1}) with (\ref{Rn}) for several 
    values of $\nu$ at time 
    $t=0.4$.}
    \label{figkp1e2m4nu}
\end{figure}

In the following we only consider the case 
$\epsilon=0.1$. The corresponding solution to the KdV equation can 
be seen in Fig.~\ref{figkdv1e2der}.
For simplicity, we shall only compare solutions for different values of $\nu$ plotted at $y=0$, \ie the region 
where the oscillations are the 
most pronounced. In Fig.~\ref{figkp1e24nu} the solution to 
the KP-I equation with initial data (\ref{kpld1}), (\ref{Rn}), is shown for several values of 
$\nu$. 
It can be seen that the case $\nu=0.01$ is very close to the KdV situation. 
Thus a continuous deformation of the KdV sector to a 
nontrivial $y$-dependence is possible, also numerically. Increasing 
$\nu$ has two effects: First we observe that the oscillations for 
$x<0$ are suppressed whereas the oscillations for $x>0$ are enhanced. 
This qualitative behavior has already been encountered above by comparing 
solutions of the KP-I and KP-II equation for the same initial data. 
Secondly, since a stronger $y$-dependence enhances the the formation of tails we observe that, the bigger $\nu$ is, the 
more mass is transferred from the initial wave pulse to  
these tails. This implies that less mass is available to form shocks in 
the dispersionless case and thus rapid oscillations for small 
$\e$ are severely damped for large $\nu$, as is clearly visible in the case $\nu = 10$.  
\begin{remark} 
In the case $\nu = 10$ the periodic boundary conditions produces very strong echoes, as it is 
clearly visible in Fig.~\ref{figkp1e2nu10}. This however is the only example considered here, 
where the echoes dominate the oscillations which we want to study. 
\end{remark}
Finally the corresponding transition in $\nu$ for the KP-II case ($\lambda=1$) starting from a situation close to KdV is 
demonstrated in Fig.~\ref{figkp1e2m4nu}. There the 
oscillations for negative $x$ are enhanced whereas the oscillations 
for positive $x$ are suppressed as $\nu$ increases. For large 
values of $\nu$ we again find that more and more mass is transferred 
to the tails, \ie pushed to $x\to\- -\infty$. 
In summary we find that the tails tend to suppress strong 
gradients, and thus shocks in the dKP equation, as well as the corresponding rapid 
oscillations in \eqref{KP} for small $\e$.


\bibliographystyle{amsplain}

\end{document}